\documentclass[10pt,a4paper]{article}
\usepackage[a4paper,top=3cm,bottom=3cm,left=2cm,right=2cm,bindingoffset=5mm]{geometry}
\usepackage[usenames, dvipsnames]{color}
\usepackage[greek,english]{babel}
\usepackage{amsmath}
\usepackage{amssymb}
\usepackage[usenames,dvipsnames]{color}
\usepackage{cancel}
\usepackage{graphicx}
\usepackage{mathrsfs}
\usepackage{bbm}
\usepackage{bm}
\usepackage{cite}
\usepackage{hyperref}

\numberwithin{equation}{section}

\hypersetup{
colorlinks=true,         
linkcolor=blue,          
citecolor=red,        
urlcolor=red            
}

\makeatletter
\newcommand{\pushright}[1]{\ifmeasuring@#1\else\omit\hfill$\displaystyle#1$\fi\ignorespaces}
\newcommand{\pushleft}[1]{\ifmeasuring@#1\else\omit$\displaystyle#1$\hfill\fi\ignorespaces}
\makeatother

\usepackage{scalerel}

\newcommand\E{{\,\scaleobj{1.3}{\bm{e}}}}

\newcommand\Eos{{\,\scaleobj{1.3}{\mathbbm{e}}}}

\newcommand\A{{\,\scaleobj{1.3}{\bm{a}}}}

\newcommand\F{{\,\scaleobj{1.3}{\bm{f}}}}

\begin{document}

\title{\vspace{-2.5cm}\bf  $\kappa$-Poincar\'e-comodules, Braided Tensor Products and Noncommutative Quantum Field Theory}

\author{{\Large Fedele Lizzi$^{a,b,c}$\footnote{fedele.lizzi@na.infn.it}, Flavio Mercati$^{a,b}$\footnote{flavio.mercati@gmail.com}}
\vspace{12pt}
\\
$^{a}$ Dipartimento di Fisica ``Ettore Pancini'',
\\
Universit\`{a} di Napoli {\sl Federico~II}, Napoli, Italy;
\vspace{6pt}
\\
$^{b}$ INFN, Sezione di Napoli,
\vspace{6pt}
\\
$^{c}$ Departament de F\'{\i}sica Qu\`antica i Astrof\'{\i}sica\\
and Institut de C\'{\i}encies del Cosmos (ICCUB),\\
Universitat de Barcelona, Barcelona, Spain.
}

\date{}

\maketitle 

\begin{abstract}
We discuss the obstruction to the construction of a multiparticle field theory on a $\kappa$-Minkowski noncommutative spacetime: the existence of multilocal functions which respect the deformed symmetries of the problem. This construction is only possible for a light-like version of the commutation relations, if one requires invariance of the tensor product algebra under the coaction of the $\kappa$-Poincar\'e group. This necessitates a \emph{braided} tensor product. We study the representations of this product, and prove that $\kappa$-Poincar\'e-invariant N-point functions belong to an Abelian subalgebra, and are therefore commutative. We use this construction to define the 2-point Whightman and Pauli--Jordan functions, which turn out to be identical to the undeformed ones. We finally outline how to construct a free scalar $\kappa$-Poincar\'e-invariant quantum field theory, and identify some open problems.
\end{abstract}

\newpage

\section{Introduction}

The $\kappa$-Minkowski spacetime \cite{Lukierski:1991ff,Lukierski:1992dt,Majid:1994cy} is a  deformation of the algebra of complex-valued functions on Minkowski spacetime, $\mathbbm{C}[\mathbbm{R}^{3,1}]$ into the noncommutative *-algebra $\mathcal{A}$, generated by the coordinate functions\footnote{See~\cite{Aschieri:2005zs,Moyalarea,Martinetti:2011mp} for another example.}
\begin{equation}\label{kappa-Minkowski_generalized}
[x^\mu , x^\nu] = \frac i \kappa  ( v^\mu x^\nu -v^\nu x^\mu) \,, \qquad \mu = 0, \dots, 3\,, \qquad (x^\mu)^\dagger = x^\mu \,,
\end{equation}
where $v^\mu $ are four arbitrary real numbers, and the $x^\mu$ operators generalize the Cartesian coordinate functions. The constant $\kappa$ has the dimensions of an inverse length, supposedly identified with (or at least related to) the Planck energy. From now on, we will work in units in which $\kappa = 1$. The above relations close a Lie algebra, known as $\mathfrak{an}(3)$, of which $\mathcal{A}$ is the universal enveloping algebra. Notice that all these algebras, for any choice of $v^\mu$, are isomorphic to each other. This can be seen by observing that the following linear redefinition of generators: $x^i \to v^0 x^i - v^i x^0$, $x^0 \to v_i x^i + \frac{1- \| \vec v \|^2}{v^0} x^0$ puts the algebra in the form:
\begin{equation}\label{kappa-Minkowski_timelike}
[x^0,x^i]=\frac{i}{\kappa}x^i \,, \qquad \left[x^i,x^j\right]=0 \,,  \qquad i,j = 1,2,3 \,,
\end{equation}
which is the original~\cite{Majid:1994cy} and best-known form of the $\kappa$-Minkowski algebra.
The algebra~(\ref{kappa-Minkowski_generalized}) is invariant under the following Hopf algebra:
\begin{equation}\label{kappaPgroup}
\begin{aligned}
 \Delta[\Lambda^\mu{}_\nu]&=\Lambda^\mu{}_\alpha\otimes\Lambda^\alpha{}_\nu, &  [\Lambda^\mu{}_\nu \Lambda^\alpha{}_\beta]&=0,
\\
 \Delta[a^\mu]&=\Lambda^\mu{}_\nu\otimes a^\nu+a^\mu\otimes\mathbbm{1}, & [\Lambda ^\mu{}_\nu , a^\gamma] &= i\left[\left( \Lambda^\mu{}_\alpha \, v^\alpha - v^\mu \right)\Lambda^\gamma{}_\nu+\left( \Lambda^\alpha{}_\nu  \tilde{g}_{\alpha\beta}  -  \tilde{g}_{\nu\beta}  \right)v^\beta g^{\mu\gamma}\right],
\\
 S[\Lambda]&=\Lambda^{-1}, ~~ S[a^\mu] = - a^\mu, & [ a^\mu , a^\nu ] &= i \left( v^\mu \, a^\nu - v^\nu \, a^\mu \right) ,
\\
 \varepsilon [\Lambda^\mu_{}\nu]&=\delta^\mu{}_\nu  , ~~
\varepsilon [a^\mu]=0 , & 
\Lambda^\mu{}_\alpha \Lambda^\nu{}_\beta g^{\alpha\beta} &= g^{\mu\nu},  ~~ \Lambda^\rho{}_\mu \Lambda^\sigma{}_\nu g_{\rho\sigma} =g_{\mu\nu}. 
\end{aligned}
\end{equation}
where $g_{\mu\nu}$ is any symmetric invertible matrix,  and $g^{\mu\nu}$ is its inverse. While the metric is usually taken to be the standard Minkowski one,  $\eta_{\mu\nu} =\text{diag} ( -1 , + 1 ,+ 1 , +1)$, other choices are possible, including some degenerate cases~\cite{Lizzi:2020tci}.  When $g_{\mu\nu} = \eta_{\mu\nu}$, this Hopf algebra (or quantum group~\cite{majid_1995}) is called $\kappa$-Poincar\'e~\cite{Lukierski:1991ff,Lukierski:1992dt,LUKIERSKI1991331,MajidBicross,Majid:1994cy,Majid:1999td,Lukierski:2015zqa,Lizzi:2019wto,Lizzi:2018qaf}. In this case the relations~(\ref{kappaPgroup}) are to be understood as the deformation of the algebra of functions on the Poincar\'e group, $\mathbbm{C}[ISO(3,1)]$ into a noncommutative Hopf algebra $\mathcal{P}_\kappa$, in which the coproduct $\Delta$, antipode $S$ and counit $\varepsilon$ are undeformed, and simply codify the Lie group structure of $ISO(3,1)$, while the commutation relations acquire a dependence on $\kappa$, and make the algebra of function nonabelian. The operators $a^\mu$ (translations) and $\Lambda^\mu{}_\nu$ (Lorentz matrices) are coordinate functions on the group, and the matrices $\Lambda^\mu{}_\nu$ leave $g^{\mu\nu}$ and its inverse invariant in the ordinary, algebraic sense expressed by the last line of the Equation above. Moreover, Equations~(\ref{kappaPgroup}) leave the commutation relations~(\ref{kappa-Minkowski_timelike}) invariant, in the sense that the following left co-action:	
\begin{equation}\label{kappa-Coaction}
x'^\mu = \Lambda^\mu{}_\nu\otimes x^\nu+a^\mu\otimes 1 \,,
\end{equation}
is an algebra homomorphism for the relations~(\ref{kappa-Minkowski_timelike}); in other words, $\kappa$-Minkowski is a $\kappa$-Poincar\'e-comodule algebra~\cite{majid_1995}. This coaction can be seen as the rule to transform a $\kappa$-Minkowski coordinate into a $\kappa$-Poincar\'e transformed one, which is an object that lives in the tensor product $\mathcal{P}_\kappa \otimes\mathcal A$, the noncommutative version of the algebra of functions on $ISO(3,1)\times\mathbbm{R}^{3,1}$.

Depending on the choice of eigenvalues of the matrix $g_{\mu\nu}$, the Hopf algebra~(\ref{kappaPgroup}) might be a quantum-group deformation of the Poincar\'e, Euclidean or $ISO(2,2)$ groups. Moreover, there are also degenerate cases in which $g_{\mu\nu}$ is not invertible, but the algebra is still well-defined, and it might, for example, correspond to a deformation of the Carroll group~\cite{Lizzi:2020tci}.  According to the particular form of  $g_{\mu\nu}$, \emph{i.e.} in which directions its eigenvectors are pointing, the coordinates $x^0$, $x^1$, $x^2$ and $x^3$ might have different nature. In the Poincar\'e case $g_{\mu\nu} = \eta_{\mu\nu}$, for example, $x^0$ is the timelike direction and $x^i$ are the spacelike ones. But any other choice is possible (and linear combinations thereof). Similarly, the vector $v^\mu$ in Eq.~(\ref{kappa-Minkowski_generalized}) could take any form, and if it is pointing in the $0$ direction, $v^\mu = \delta^\mu_0$, then the commutation relations reduce to~(\ref{kappa-Minkowski_timelike}), in which $x^0$ is the only noncommuting coordinate. In all other cases, the direction of $v^\mu$ indicates which linear combination of $x^\mu$ coordinates is the noncommuting one. Of course, one can act on the generators $x^\mu$ with any (commutative) linear transformation, and obtain an algebra with a different  $v^\mu$ vector, invariant under a quantum group~(\ref{kappaPgroup}) with a different matrix $g^{\mu\nu}$. In the end, in the case that $g^{\mu\nu}$ is invertible, what counts is the relative orientation of $v^\mu$ with respect to the eigenvectors of $g^{\mu\nu}$. In the degenerate cases things are more complicated. For a complete treatment of all the physically-inequivalent models, and the corresponding momentum spaces, see~\cite{Lizzi:2020tci}.

Once we have a generalization of the algebra of functions on a manifold, the natural context to look for physical applications of the model is field theory, whose basic ontology is that of fields, which are multiplets of functions on the spacetime manifold. Classical (in the sense of unquantized, $\hbar \to 0$ limit) noncommutative models based on action functionals and equations of motion are fairly well understood~\cite{Lukierski:1993wxa,KowalskiGlikman:2002jr,Agostini:2002yd,Agostini:2002de,KowalskiGlikman:2002ft,KowalskiGlikman:2003we,Agostini:2005mf,Amelino-Camelia2007,AmelinoCamelia:2007rn,Carmona:2011wc,Carmona:2012un,Gravityquantumspacetime,Gubitosi:2013rna,Mercati:2011aa,Lukierski:2015zqa,Meljanac:2016jwk,Loret:2016jrg,Lukierski:2016vah,Mercati:2018fba}. There is, however, no current agreement in the literature on the correct formulation of noncommutative Quantum Field Theory (QFT), although there is a sizeable literature on the subject~\cite{Kosinski1999,Kosinski2001,Kosinski:2003xx,Arzano2007,Daszkiewicz:2007az,Freidel:2007hk,Arzano:2009ci,Arzano:2017uuh,Juric2015}. Recently, there has been a resurgence in interest for QFT $\kappa$-Minkowski~\cite{Mathieu2020,Poulain2018a,Poulain2018,Juric2018,Arzano2018,Mercati:2018hlc,Mercati:2018ruw,Juric2015}, and perhaps the most important difference between approaches regards the basic ontology. Most approaches are based on a commutative algebra of functions, over which a non-local ``star'' product, involving an infinite number of derivatives of the fields, is defined. This star product provides a representation of the basic commutation relations~(\ref{kappa-Minkowski_timelike}) or~(\ref{kappa-Minkowski_generalized}), and the theory is treated as a commutative-but-nonlocal QFT, defined, for example, through a regular path integral. Assuming such an ontology might be problematic from the operational point of view~\cite{Huggett:2020kok}, and it is not clear whether $\kappa$-Poincar\'e symmetries can be implemented as symmetries of the theory. But most importantly, such an ontology naturally leads one to define the QFT in terms of ``commutative'' N-point functions (defined, \emph{e.g.} through the functional derivatives of a partition function with respect to the commutative fields) that do not address the issue of multilocal functions, which we describe in the following.

In this paper we want to attack the main obstruction that prevented the full development of a QFT  based on a truly noncommutative ontology: the fact that, in order to work with QFTs, it is necessary to have a good notion of multilocal functions, because the theory is entirely determined by its N-point functions. From an algebraic point of view, we would like to have, to begin, a notion of ``function of two points''. This is a function on the Cartesian product of two copies of Minkowski space, $\mathbbm{R}^{3,1} \times \mathbbm{R}^{3,1}$. The commutative algebra of such functions is $\mathbbm{C}[\mathbbm{R}^{3,1} \times \mathbbm{R}^{3,1}]$,  which, under the canonical isomorphism, can be identified with the tensor product algebra $\mathbbm{C}[\mathbbm{R}^{3,1} ] \otimes \mathbbm{C}[ \mathbbm{R}^{3,1}]$, which is canonically defined as generated by the coordinate functions:
\begin{equation}\label{TensorProductAlgebraGenerators}
x_1^\mu = x^\mu \otimes 1 \,, \qquad x_2^\mu = 1 \otimes x^\mu  \,, 
\end{equation}
with the identity $1^{\otimes 2} = 1 \otimes 1$, and the product is simply $x^\mu_1 x^\nu_2 = x^\nu_2 x^\mu_1$.
So, in the noncommutative setting, it appears natural to refer to the tensor product algebra $\mathcal{A}^{\otimes 2}$ generated by~(\ref{TensorProductAlgebraGenerators}), where
\begin{equation} \label{TensorProductCommutators}
[x_1^\mu , x_1^\nu] =i  ( v^\mu x_1^\nu -v^\nu x_1^\mu) \,, \qquad [x_2^\mu , x_2^\nu] =i  ( v^\mu x_2^\nu -v^\nu x_2^\mu) \,,
\qquad 
[x_1^\mu , x_2^\nu] = 0 \,.
\end{equation}
This algebra is a good Lie algebra (it satisfies the Jacobi rules), and gives rise to a perfecly legitimate universal enveloping algebra. It also makes sense that the coordinates $x^\mu_1$ and $x^\mu_2$ are the operators that generalize to the noncommutative settings the coordinates of point 1 and point 2, which are distinct points which we should be able to choose independently.\footnote{With ``choosing a point'', in the noncommutative setting, we mean choosing a state on the algebra, which can provide a degree of localization. In fact, classical points can be described through the commutative algebra of functions on a manifold as limits of functions peaked around a choice of coordinates (\emph{e.g.} Gaussians), which tend to a Dirac delta. In the noncommutative setting this limit is unattainable except for special points (\emph{e.g.} the time axis, \cite{Lizzi:2018qaf}), because of uncertainty relations. However, one can introduce a notion of ``fuzzy points'', corresponding to maximally-localized states~\cite{Mercati:2018hlc,Mercati:2018ruw,Lizzi:2018qaf}.} Since, by construction, the states on the tensor product algebra allow us to localize $x^\mu_1$ and $x^\mu_2$ independently around arbitrary classical coordinates without interference of the state of one point on the other, we could be quite satisfied with this formulation.
However, there is a big problem: extending the $\kappa$-Poincar\'e coaction~(\ref{kappa-Coaction}) to $\mathcal A^{\otimes 2}$ in the canonical way:
\begin{equation}
\begin{aligned}
{x'}_1^\mu =  \Lambda^\mu{}_\nu\otimes x_1^\nu + a^\mu\otimes 1^{\otimes 2}
=  \Lambda^\mu{}_\nu\otimes x^\nu \otimes 1 + a^\mu\otimes 1 \otimes 1 \,,
\\
{x'}_2^\mu =  \Lambda^\mu{}_\nu\otimes x_2^\nu + a^\mu\otimes 1^{\otimes 2}
=  \Lambda^\mu{}_\nu\otimes 1 \otimes x^\nu  + a^\mu \otimes 1 \otimes 1 \,,
\end{aligned}
\end{equation}
the algebra~(\ref{TensorProductCommutators}) is not left invariant by it. In technical terms, (\ref{TensorProductCommutators}) is not a $\kappa$-Poincar\'e-comodule. Specifically, it is the commutation relations between $x^\mu_1$ and $x^\mu_2$ that are not covariant. In fact, 
\begin{equation}\label{NoncovarianceOfTensorproduct}
[{x'}_1^\mu , {x'}_2^\nu] = 
[\Lambda^\mu{}_\rho,a^\nu]\otimes \left( x_1^\rho -  x_2^\rho \right) + [a^\mu,a^\nu]\otimes 1^{\otimes 2} \neq 0 \,.
\end{equation}

There is a way out of this problem: relax the commutativity of the two sides of the tensor product algebra, $[x_1^\mu , x_2^\nu] =0$, in order to make Eq.~(\ref{NoncovarianceOfTensorproduct}) covariant.  The structure we end up dealing with is a ``braided tensor product'', first introduced by Majid in the 90's~\cite{majid1995algebras,majid_1995}. A similar concept has been used in~\cite{Oeckl:2000eg,Wess:2003da,Chaichian:2004za,Koch:2004ud,Aschieri:2005zs,Fiore:2007vg,Fiore:2007zz} to properly define QFT on the Moyal/canonical spacetime.  In~\cite{Daszkiewicz:2007az} the necessity to extend the $\kappa$-Minkowski algebra to multiple points in a nontrivial way was recognized.  The novelty in our work is that we require that the proposed solution of the problem provides a coherent $\kappa$-Poincar\'e comodule. In the following, we will address the issue and find the conditions under which it can be solved.

A related alternative, which we will not pursue here, would be to enforce the symmetry via a Drinfeld twist, and coherently generate a deformed tensor product, deformed star product and other structures, along the lines of~\cite{Aschieri:2007sq}. Twists for $\kappa$-Minkowski symmetries have been studied, they are not exempt from problems~\cite{Borowiec:2013gca}, a recent review, with references, is~\cite{Juric:2015aza}.

\section{The braided tensor product algebra}

Let us first consider the algebra of two points. We have to request that it closes two $\kappa$-Minkowski~(\ref{kappa-Minkowski_generalized}) subalgebras:
\begin{equation}
\begin{aligned}
\left[x_1^{\mu },x_1^{\nu }\right] = i \left[ v^{\mu } x_1^{\nu } - v^{\nu } x_1^{\mu }  \right] \,,
\qquad
\left[x_2^{\mu },x_2^{\nu }\right] = i \left[ v^{\mu } x_2^{\nu } - v^{\nu } x_2^{\mu }  \right] \,,
\end{aligned}
\end{equation}
with yet-to-be-determined cross-commutators:
\begin{equation}\label{Cross-Commutators_1}
\left[x_1^{\mu },x_2^{\nu }\right] = i f^{\mu\nu}(x_1,x_2,v)\,,
\end{equation}
and it should form a left-comodule under the following left coaction:
\begin{equation}\label{kappaCoaction}
x_a'^\mu = \Lambda^\mu{}_\nu  x_a^\nu +  a^\mu   \,, ~~~ a =1,2 \,,
\end{equation}
of the $\kappa$-Poincar\'e group~(\ref{kappaPgroup}). Finally, we have to request that the commutators~(\ref{Cross-Commutators_1}) satisfy the Jacobi rules.
In addition to these definitory requests, we can make a few reasonable assumptions: the function $f^{\mu\nu}(x_1,x_2,v)$ should go to zero when $v^\mu \to 0$, and we can assume it is polynomial in $x^\mu_a$. Under this ansatz,
we can expand it in powers of $v^\mu$:
\begin{equation}\label{Cross-Commutators_2}
\left[x_1^{\mu },x_2^{\nu }\right] = i  a^{\mu\nu}_{\rho\sigma} v^\rho v^\sigma + i v^\rho \left(b^{\mu\nu}_{\rho\sigma}  x_1^\sigma + c^{\mu\nu}_{\rho\sigma}  x_2^\sigma \right)  \,,
\end{equation}
where $a^{\mu\nu}_{\rho\sigma}$, $b^{\mu\nu}_{\rho\sigma}$ and $c^{\mu\nu}_{\rho\sigma} $ are numbers.
Imposing the comodule condition on this commutator:
\begin{equation}
\begin{aligned}
\left[x_1'^{\mu },x_2'^{\nu }\right] =& i  a^{\mu\nu}_{\rho\sigma} v^\rho v^\sigma + i v^\rho \left(b^{\mu\nu}_{\rho\sigma}  x_1'^\sigma + c^{\mu\nu}_{\rho\sigma}  x_2'^\sigma \right)  \,,
\\
\Lambda^\mu{}_\rho \Lambda^\nu{}_\sigma [ x_1^\rho ,   x_2^\sigma]
+
\left[ \Lambda^\mu{}_\rho, a^\nu   \right]  x_1^\rho 
+
\left[   a^\mu  , \Lambda^\nu{}_\sigma   \right] x_2^\sigma 
+
\left[  a^\mu  ,  a^\nu   \right]
 =& i  a^{\mu\nu}_{\rho\sigma} v^\rho v^\sigma + i v^\rho \left(b^{\mu\nu}_{\rho\sigma} \Lambda^\sigma{}_\lambda x_1^\lambda + c^{\mu\nu}_{\rho\sigma}  \Lambda^\sigma{}_\lambda x_2^\lambda \right) 
 \\&
 + i v^\rho \left(b^{\mu\nu}_{\rho\sigma}  a^\sigma + c^{\mu\nu}_{\rho\sigma}  a^\sigma \right)  \,,
\\
 i  \Lambda^\mu{}_\rho \Lambda^\nu{}_\sigma  a^{\rho\sigma}_{\lambda\tau} v^\lambda v^\tau
  + i \Lambda^\mu{}_\rho \Lambda^\nu{}_\sigma   v^\lambda \left(b^{\rho\sigma}_{\lambda\tau}  x_1^\tau + c^{\rho\sigma}_{\lambda\tau}  x_2^\tau \right)
 =& i  a^{\mu\nu}_{\rho\sigma} v^\rho v^\sigma + i v^\rho \left(b^{\mu\nu}_{\rho\sigma} \Lambda^\sigma{}_\lambda x_1^\lambda + c^{\mu\nu}_{\rho\sigma}  \Lambda^\sigma{}_\lambda x_2^\lambda \right) 
 \\
 +\left[ \Lambda^\mu{}_\rho, a^\nu   \right]  x_1^\rho 
+
\left[   a^\mu  , \Lambda^\nu{}_\sigma   \right] x_2^\sigma 
+
\left[  a^\mu  ,  a^\nu   \right] ~~~~~&
 + i v^\rho \left(b^{\mu\nu}_{\rho\sigma}  a^\sigma + c^{\mu\nu}_{\rho\sigma}  a^\sigma \right)  \,.
 \end{aligned}
\end{equation}
The different powers of $v^\mu$ in the above equation have to vanish separately. The quadratic term gives
\begin{equation}
  \Lambda^\mu{}_\rho \Lambda^\nu{}_\sigma  a^{\rho\sigma}_{\lambda\tau} v^\lambda v^\tau
  =  a^{\mu\nu}_{\rho\sigma} v^\rho v^\sigma \,,
\end{equation}
which cannot be solved if $a^{\mu\nu}_{\rho\sigma}  \neq 0$, so we have to put it to zero. We then split the terms that are linear in $x_a^\mu$ from the one that does not depend on it, which reads:
\begin{equation}
\left[  a^\mu  ,  a^\nu   \right] \equiv i \left( v^\mu \, a^\nu - v^\nu \, a^\mu \right) =   i v^\rho \left(b^{\mu\nu}_{\rho\sigma}  + c^{\mu\nu}_{\rho\sigma} \right)  a^\sigma  \,.
\end{equation}
This is solved by
\begin{equation}\label{Eq_comodule0}
b^{\mu\nu}_{\rho\sigma}  + c^{\mu\nu}_{\rho\sigma} =  \delta^\mu{}_\rho \delta^\nu{}_\sigma - \delta^\nu{}_\rho \delta^\mu{}_\sigma \,.
\end{equation}
The two terms that are linear in $x^\mu_1$ and, respectively, in $x^\mu_2$ vanish iff
\begin{equation}
  i \Lambda^\mu{}_\rho \Lambda^\nu{}_\sigma   v^\lambda b^{\rho\sigma}_{\lambda\tau}  
  +\left[ \Lambda^\mu{}_\rho, a^\nu   \right]  \delta^\rho{}_\tau 
= i v^\rho b^{\mu\nu}_{\rho\sigma} \Lambda^\sigma{}_\tau \,,
\qquad
  i \Lambda^\mu{}_\rho \Lambda^\nu{}_\sigma   v^\lambda  c^{\rho\sigma}_{\lambda\tau}   
+
\left[   a^\mu  , \Lambda^\nu{}_\sigma   \right] \delta^\sigma{}_\tau 
= i v^\rho c^{\mu\nu}_{\rho\sigma}  \Lambda^\sigma{}_\tau \,.
\end{equation}
Using the $\kappa$-Poincar\'e relations $ [\Lambda ^\mu{}_\nu , a^\gamma] = i\left[\left( \Lambda^\mu{}_\alpha \, v^\alpha - v^\mu \right)\Lambda^\gamma{}_\nu+\left( \Lambda^\alpha{}_\nu  \eta_{\alpha\beta}  -  \eta_{\nu\beta}  \right)v^\beta \eta^{\mu\gamma}\right]$ we can write these two equations as
%
%
%
%
\begin{equation}\label{Eq_comodule}
\begin{aligned}
\Lambda^\mu{}_\rho \Lambda^\nu{}_\sigma   v^\lambda b^{\rho\sigma}_{\lambda\tau}  
 - v^\rho b^{\mu\nu}_{\rho\sigma} \Lambda^\sigma{}_\tau  +\left[\left( \Lambda^\mu{}_\alpha \, v^\alpha - v^\mu \right)\Lambda^\nu{}_\rho+\left( \Lambda^\alpha{}_\rho  \eta_{\alpha\beta}  -  \eta_{\rho\beta}  \right)v^\beta \eta^{\mu\nu}\right]  \delta^\rho{}_\tau
= 0\,,
\\
\Lambda^\mu{}_\rho \Lambda^\nu{}_\sigma   v^\lambda  c^{\rho\sigma}_{\lambda\tau}   
- v^\rho c^{\mu\nu}_{\rho\sigma}  \Lambda^\sigma{}_\tau- \left[\left( \Lambda^\nu{}_\alpha \, v^\alpha - v^\nu \right)\Lambda^\mu{}_\sigma+\left( \Lambda^\alpha{}_\sigma  \eta_{\alpha\beta}  -  \eta_{\sigma\beta}  \right)v^\beta \eta^{\nu\mu}\right]
 \delta^\sigma{}_\tau 
=  0 \,.
\end{aligned}
\end{equation}
To solve these equations, we should recall that $\Lambda^\mu{}_\nu$ is an $SO(3,1)$ matrix, which can therefore be expanded in powers of an antisymmetric matrix $\epsilon_{\alpha\beta}$ as
\begin{equation}
\Lambda^\mu{}_\nu = \delta^\mu{}_\nu + \epsilon_{\rho\nu} \eta_{\rho\mu} + \mathcal{O}(\epsilon^2)\,.
\end{equation}
Eq.~(\ref{Eq_comodule}) reads, at first order in  $\epsilon^{\alpha\beta}$:
\begin{equation}\label{Eq_comodule2}
\begin{aligned}
\epsilon_{\alpha\beta} v^\lambda \left(  \eta^{\mu \alpha} b^{\beta\nu}_{\lambda \tau} + \eta^{\nu \alpha} b^{\mu \beta}_{\lambda \tau} - \delta^\beta{}_\tau \eta^{\rho \alpha} b^{\mu\nu}_{\lambda \rho}  + \delta^\beta{}_\lambda \delta^\nu{}_\tau \eta^{\mu \alpha} + \delta^\beta{}_\tau \delta^\alpha{}_\lambda \eta^{\mu\nu}   \right) =  0 \,,
\\
\epsilon_{\alpha\beta} v^\lambda \left(  \eta^{\mu \alpha} c^{\beta\nu}_{\lambda \tau} + \eta^{\nu \alpha} c^{\mu \beta}_{\lambda \tau} - \delta^\beta{}_\tau \eta^{\rho \alpha} c^{\mu\nu}_{\lambda \rho}  - \delta^\beta{}_\lambda \delta^\nu{}_\tau \eta^{\mu \alpha} - \delta^\beta{}_\tau \delta^\alpha{}_\lambda \eta^{\mu\nu}   \right) =  0 \,,
\end{aligned}
\end{equation}                                                                                                                                                        
which are equivalent to
\begin{equation}\label{Eq_comodule3}
\begin{aligned}
 \eta^{\mu [\alpha} b^{\beta]\nu}_{\lambda \tau} +  b^{\mu [\beta}_{\lambda \tau} \eta^{\alpha]\nu} - \delta^{[\beta}{}_\tau \eta^{\alpha]\rho} b^{\mu\nu}_{\lambda \rho} 
  +
 \eta^{\mu [\alpha}   \delta^{\beta]}{}_\lambda \delta^\nu{}_\tau  + \delta^{[\beta}{}_\tau \delta^{\alpha]}{}_\lambda \eta^{\mu\nu}   =  0 \,,
\\
 \eta^{\mu [\alpha} c^{\beta]\nu}_{\lambda \tau} +  c^{\mu [\beta}_{\lambda \tau} \eta^{\alpha]\nu} - \delta^{[\beta}{}_\tau \eta^{\alpha]\rho} c^{\mu\nu}_{\lambda \rho} 
-
 \eta^{\mu [\alpha}   \delta^{\beta]}{}_\lambda \delta^\nu{}_\tau  - \delta^{[\beta}{}_\tau \delta^{\alpha]}{}_\lambda \eta^{\mu\nu}   =  0 \,.
\end{aligned}
\end{equation}                                                                                                                                                        
The two equations above are satisfied by
\begin{equation}\label{Comodule_sol}
b^{\mu\nu}_{\rho\sigma} = \delta^\mu{}_\rho \delta^\nu{}_\sigma - \eta^{\mu\nu} \eta_{\rho\sigma} \,,
\qquad
c^{\mu\nu}_{\rho\sigma} = -\delta^\nu{}_\rho \delta^\mu{}_\sigma + \eta^{\mu\nu} \eta_{\rho\sigma} \,,
\end{equation}
which satisfies also Eq.~(\ref{Eq_comodule0}). A quick calculation reveals that this perturbative solution is exact at all orders in $\epsilon_{\mu\nu}$. In fact, replacing~(\ref{Comodule_sol}) into Eq.~(\ref{Eq_comodule}) the two equations reduce to $v^\lambda \left( \eta^{\mu\nu} - \Lambda^\mu{}_\rho \Lambda^\nu{}_\sigma \eta^{\rho\sigma} \right) = 0 $, which is of course satisfied as long as $\Lambda^\mu{}_\nu  \in SO(3,1)$.

We then found a general solution of the comodule problem:
\begin{equation}
\left[x_1^{\mu },x_2^{\nu }\right] = i \left[ v^{\mu } x_1^{\nu } - v^{\nu } x_2^{\mu }  -\eta^{\mu \nu } \eta_{\rho \sigma } v^{\rho } \left(x_1^{\sigma }- x_2^{\sigma }\right) \right] \,.
\end{equation}
Notice now that the above commutators can be written in the following form:
\begin{equation}
\left[x_a^{\mu },x_b^{\nu }\right] = i \left[ v^{\mu } x_a^{\nu } - v^{\nu } x_b^{\mu }  -\eta^{\mu \nu } \eta_{\rho \sigma } v^{\rho } \left(x_a^{\sigma }- x_b^{\sigma }\right) \right] \,,
\end{equation}
which reduce to the usual (generalized) $\kappa$-Minkowski commutators when $a=b$:
\begin{equation}
\left[x_a^{\mu },x_a^{\nu }\right] = i \left( v^{\mu } x_a^{\nu } - v^{\nu } x_a^{\mu } \right) \,,
\end{equation}
and, moreover, remain consistent even if we let the indices $a,b$ run on an arbitrarily large set of labels. We have a comodule regardless of the number of points we are considering.

In order to have a proper (associative) comodule algebra, our commutators need to satisfy also the 
Jacobi rules:
\begin{equation}
[x_a^\mu , [x_b^\nu , x_c^\rho]]  + [x_b^\nu [ x_c^\rho, x_a^\mu]] + [x_c^\rho, [ x_a^\mu , x_b^\nu ]] = 0 \,.
\end{equation}
A straightforward, but tedious, calculation reveals that
\begin{equation}
[x_a^\mu , [x_b^\nu , x_c^\rho]]  + [x_b^\nu [ x_c^\rho, x_a^\mu]] + [x_c^\rho, [ x_a^\mu , x_b^\nu ]] = 
-   v^\alpha v_\alpha  \left[ \eta^{\nu \rho } (x_c^{\mu }  -  x_b^{\mu })  +  \eta^{\rho \mu } (x_a^{\nu }  -  x_c^{\nu }) +   \eta^{\mu \nu }    (x_b^{\rho }  -  x_a^{\rho }) \right] \,,
\end{equation}
and the only way that the right-hand side can vanish is that $ v^\alpha v_\alpha=0$.

We obtained a significant result: the only $\kappa$-Minkowski-like algebra that admits a braided tensor product construction as a $\kappa$-Poincar\'e comodule is the \emph{lightlike} one, in which the deformation parameters $v^\mu$ form a lightlike vector. Our result is coherent with the one found by Juri\'c, Meljanac and Pikuti\'c~\cite{Juric:2015aza} using a Drinfeld twist. They too obtained that a covariant deformation of the tensor product can only be obtained for the lightlike case. This particular choice for the vector $v$ is remarkable for several other reasons, and the result we just derived makes it the only viable algebra, in the $\kappa$-Minkowski family, to construct a well-defined quantum field theory. 
%


\section{Representation of the braided $\kappa$-Minkowski algebra}

Let us review the results obtained so far. The following algebra, which we will call $\mathcal{A}^{\underline{\otimes}N}$, generated by the identity together with $4N$ generators $x^\mu_a$, $a=1,\dots, N$:
\begin{equation}\label{Braided-kappa-Minkowski}
\left[x_a^{\mu },x_b^{\nu }\right] = i \left[ v^{\mu } x_a^{\nu } - v^{\nu } x_b^{\mu }  -\eta^{\mu \nu } \eta_{\rho \sigma } v^{\rho } \left(x_a^{\sigma }- x_b^{\sigma }\right) \right] \,, \qquad x^\mu_a \in \mathcal{A}^{\underline{\otimes}N} \,,
\end{equation}
is a left comodule for the $\kappa$-Poincar\'e group:
\begin{equation}\label{kappaPoincareGroup}
\begin{gathered}
\left[ a^\mu , a^\nu \right] =  i \left( v^\mu a^\nu - v^\nu a^\mu \right) \,, \qquad [ \Lambda^\mu{}_\nu ,  \Lambda^\rho{}_\sigma ] = 0 \,,
\\
\left[a^{\alpha },\Lambda ^{\mu }{}_{\nu }\right] = i \left[  \left( v^{\beta } \Lambda ^{\mu }{}_{\beta }-v^{\mu } \right)  \Lambda ^{\alpha }{}_{\nu } +
  \left( \Lambda ^{\beta }{}_{\nu } \eta_{\beta \rho }-\eta_{\nu \rho }\right) v^{\rho } \eta^{\alpha \mu } \right] \,,
\end{gathered}
\end{equation}
with respect to the coaction $x_a'^\mu = \Lambda^\mu{}_\nu  x_a^\nu +  a^\mu $, if the vector $v^\mu$ is light-like ($v^\mu v^\nu \eta_{\mu\nu}=0$.

We now proceed to study the representations of the algebra~(\ref{Braided-kappa-Minkowski}). To start, notice that the relative positions:
\begin{equation}
\Delta x_{ab}^\mu = x_a^\mu - x_b^\mu \,,
\end{equation}
close an Abelian subalgebra:
\begin{equation}
[ \Delta x^\mu_{ab}, \Delta x^\nu_{cd} ] = 0 \qquad \forall ~ a,b,c,d \,.
\end{equation}
These however are wildly redundant. If we are interested in identifying the maximal abelian subalgebra we should introduce the `centre of mass' coordinates:
\begin{equation}
x_\text{cm}^\mu = \frac 1 N \sum_{a=1}^N x_a^\mu \,,
\qquad
y^\mu_a = x^\mu_a -  x_\text{cm}^\mu \,,
\end{equation}
then it is easy to show that
\begin{equation}
[ y^\mu_{a}, y^\nu_{b} ] = 0 \qquad \forall ~ a,b \,.
\end{equation}
The $y^\mu_a$ are $4N$ variables, but $4$ of these are redundant, because they satisfy the linear relation $\sum_{a=1}^N y_a^\mu  = 0$. So we have identified a $4(N-1)$-dimensional Abelian subalgebra. What about the remaining four variables, $x_\text{cm}^\mu$?
Their commutators with $ y_a^\nu $ give a linear combination of  $ y_a^\nu$, and they close a $\kappa$-Minkowski subalgebra with each other:
\begin{equation}\label{XcmAlgebra}
[ x_\text{cm}^\mu , y_a^\nu ] = i \left( \eta^{\mu\nu} \eta_{\rho\sigma} v^\rho y^\sigma_a   - v^\nu y^\mu_a \right) \,, ~~ 
[x_\text{cm}^\mu , x_\text{cm}^\nu] = i \left( v^{\mu } x_\text{cm}^{\nu } - v^{\nu } x_\text{cm}^{\mu } \right)  \,, 
\end{equation}
however  the component of $x_\text{cm}^\mu$ along $v^\mu$ commutes with all the  $ y_a^\nu $:
\begin{equation}
w = \eta_{\mu\nu} v^\mu x_\text{cm}^\nu ~~ \Rightarrow ~~ [w , y^\mu_a ] = 0 \,, ~~ 
[x_\text{cm}^\mu , w ] = i \, v^{\mu } w   \,,
\end{equation}
we identified a $(4N -3)$-dimensional Abelian subalgebra, generated by $y_a^\mu$ and $w$, while the three components of $x_\text{cm}^\mu$ perpendicular to $v^\mu$ are irreducibly noncommutative.

Without loss of generality, we may assume $v^\mu = (1,1,0,0)$ (taking $v^\mu$ lightlike necessarily selects a special spatial direction). Then the only noncommutative coordinates are $z = x_\text{cm}^0 + x_\text{cm}^1$, $u = x_\text{cm}^2$ and $v = x_\text{cm}^3$, and the braided tensor product algebra is described by the following relations: 
\begin{equation}
\begin{aligned}
&[ y^\mu_{a}, y^\nu_{b} ] = [ y^\mu_{a}, w ] =  0 \,, \qquad \sum_{a=1}^N y^\mu_a = 0\,,
\\
&[z , u] = 2 i \, u \,, \qquad  [z , v] = 2 i \, v \,, \qquad  [z,w]= 2 i \, w \,, 
\qquad [u,w] = [v,w]= [u,v] =  0 \,,
\\
&
\\
&[ u, y_a^\nu ] = i \left( \eta^{2\nu} ( y^1_a - y^0_a )   - \left(\delta^\nu_0 + \delta^\nu_1 \right)  y^2_a \right) \,, ~~
[ v, y_a^\nu ] = i \left( \eta^{3\nu} ( y^1_a - y^0_a )   - \left(\delta^\nu_0 + \delta^\nu_1 \right) y^3_a \right) \,,
\\
&
\\
& [ z , y_a^\nu ]  = i \left( ( \delta^\nu_1 - \delta^\nu_0 ) ( y^1_a - y^0_a )   - \left(\delta^\nu_0 + \delta^\nu_1 \right)  (y^0_a + y^1_a) \right) \,.
\end{aligned}
\end{equation}
We can write a representation of the above algebra. The operators $y^\mu_a$, $a=1,\dots,N-1$ and $w$ are multiplicative with real spectrum, while the $N$-th coordinate is the a linear combination of the others: $y^\mu_N = - \sum_{a=1}^{N-1} y^\mu_a$. Finally, $u$,  $v$ and $z$ can be represented as  the following Hermitian operators:
\begin{equation}\label{RepresentationBraidedAlgebra}
\begin{aligned}
\hat u \, \psi(y^\mu_1,\dots,y^\mu_{N-1},w) &= i \sum_{a=1}^{N-1} \left(   y^1_a \frac{\partial }{\partial y^2_a}  -  y^2_a  \frac{\partial }{\partial y^1_a} - y^0_a \frac{\partial }{\partial y^2_a}    -   y^2_a  \frac{\partial }{\partial y^0_a} \right) \psi(y^\mu_1,\dots,y^\mu_{N-1},w) \,,
\\
\hat v \, \psi(y^\mu_1,\dots,y^\mu_{N-1},w) &= i \sum_{a=1}^{N-1} \left(   y^1_a \frac{\partial }{\partial y^3_a}     -   y^3_a  \frac{\partial }{\partial y^0_a} - y^0_a \frac{\partial }{\partial y^3_a}  -   y^3_a  \frac{\partial }{\partial y^1_a} \right) \psi(y^\mu_1,\dots,y^\mu_{N-1},w)\,,
\\
\hat z \, \psi(y^\mu_1,\dots,y^\mu_{N-1},w) &= - 2 i \left( y^0_a \frac{\partial }{\partial y^1_a} + y^1_a \frac{\partial }{\partial y^0_a} - w \frac{\partial}{\partial w} - \frac 1 2 \right)   \psi(y^\mu_1,\dots,y^\mu_{N-1},w)  \,.
\end{aligned}
\end{equation}
Introducing the operators that generate the simultaneous Lorentz transformations of the $N-1$ coordinates~$y^\mu_a$:
\begin{equation}
M^{\mu\nu} = i \, \sum_{a=1}^{N-1} \left(  y^\mu_a \eta^{\nu\rho} \frac{\partial }{\partial y^\rho_a}  -  y^\nu_a \eta^{\mu\rho} \frac{\partial }{\partial y^\rho_a}  \right) \,,
\end{equation}
we notice that we are representing $u$, $v$ and $z$ as:
\begin{equation}
u = M^{12} - M^{02} \,, \qquad v = M^{13} - M^{03} \,, \qquad z = 2 M^{10} + 2 \, i \, w \frac{\partial}{\partial w} + i \,,
\end{equation}
which reproduce the algebra $[z,u]=iu$, $[z,v] = iv$, $[u,v]=0$, as can be immediately verified by using the Lorentz algebra commutators $  [M^{\mu \nu },M^{\rho \sigma }] = i \left (\eta^{\nu \rho }M^{\mu \sigma } - \eta^{\mu \rho }M^{\nu \sigma } - \eta^{\nu \sigma }M^{\mu \rho }
+ \eta^{\mu \sigma }M^{\nu \rho } \right )$. 

It is not the first time that the $\kappa$-Minkowski algebra is represented as linear combinations of Lorentz generators, see for example~\cite{Blaut2004,Lizzi:2020tci}. Our braided algebra admits a representation in terms of Lorentz generators acting on the space of $4(N-1)$ spacetime points, and a dilatation operator on the real line of $w$.

Accordingly, the natural Hilbert space for the representation~(\ref{RepresentationBraidedAlgebra}) is $L^2(\mathbbm{R}^{4N-3})$, with inner product:
\begin{equation}
\langle \varphi | \psi \rangle = \int d^4 y_1 \dots d^4 y_{4N} d w  \, \bar{\varphi}(y^\mu_1,\dots,y^\mu_{N-1},w) \psi(y^\mu_1,\dots,y^\mu_{N-1},w) \,,
\end{equation}
and all our operators are self-adjoint with respect to this Hilbert space.

\section{$\kappa$-Poincar\'e-invariant Quantum Field Theory}

We will now lay the ground for a consistent construction of a QFT on the $\kappa$-Minkowski noncommutative spacetime. The first step is to define what we mean withQFT  in this context. As is well known, a QFT on a commutative spacetime (in particular Minkowski) is entirely defined in terms of all $N$-point functions~\cite{Peskin:1995ev}. We can import this definition into our noncommutative setting, however now the $N$-point functions have to be replaced with elements of our braided $N$-point algebra, which are noncommutative operators. However, the $\kappa$-Poincar\'e invariance that characterizes our theory comes to our aid. It turns out that all $\kappa$-Poincar\'e invariant elements of our braided algebra (as $N$-point functions should be) are elements of the abelian subalgebra of coordinate separations $x^\mu_a - x^\mu_b$ (or the center-of-mass coordinates $y^\mu_a$). As operators, therefore, they can all be simultaneously localized arbitrarily well, and they can be effectively treated as \emph{bona fide} commutative functions.

It is clear that all the commutative Poincar\'e-invariant polynomials remain invariant under the coaction~(\ref{kappaCoaction}). These are the functions of the (squared) proper distances:
\begin{equation}
\eta_{\mu\nu} (x_a'^\mu - x_b'^\mu)(x_a'^\nu - x_b'^\nu)= \eta_{\mu\nu} (x_a^\mu - x_b^\mu)(x_a^\nu - x_b^\nu) \qquad \forall a,b = 1,\dots ,N\,.
\end{equation}
It would be interesting to check whether these are the \emph{only} Poincar\'e-invariant polynomials in the noncommutative case, however we do not have a proof of this at the moment.
If we focus on functions that can be Fourier transformed (which is what we are interested in, if we want to define the $N$-point functions), in the commutative case one can see that
\begin{equation}
f(x'^\mu_a) = \int d^4 k^1 \dots d^4 k^N \tilde f(k_\mu^a) e^{i \sum_{a=1}^N k^a_\mu  \Lambda^\mu{}_\nu x^\nu_a} e^{i \sum_{a=1}^N k^a_\mu a^\mu } \,,
\end{equation}
is equal to $f(x^\mu_a)$ only if $\tilde f(k^a_\mu) \propto \delta^{4}\left(\sum_{a=1}^N k^a_\mu \right)$ (translation invariance), and $\tilde f( \Lambda_\nu{}^\mu k^a_\mu  ) = \tilde f(k_\nu)$ (Lorentz invariance). 

In the noncommutative case, the coordinate algebra is replaced by a Lie algebra~(\ref{Braided-kappa-Minkowski}). Therefore plane waves, \emph{i.e.} exponentials of the generators, are replaced by Lie group elements, and Fourier transforms admit a definition in terms of a group average~\cite{Mercati:2018hlc,Lizzi:2018qaf}. We can represent a generic group element once we choose a factorization, \emph{i.e.} an ordering choice. For example:
\begin{equation}
e^{i k^1_\mu  x^\mu_1 } \dots e^{i k^N_\mu  x^\mu_N}
\end{equation}
covers all group elements, upon  varying $k^1_a$ over all of $\mathbbm{R}^{4N}$. Once we introduced this ordering prescription, all Fourier-transformable functions can be represented as a linear combination of group elements:
\begin{equation} \label{NCFourier}
f(x^\mu_a) =\int d^4 k^1 \dots d^4 k^N \,  \tilde f(k_\mu^a) \,e^{i k^1_\mu  x^\mu_1 } \dots e^{i k^N_\mu  x^\mu_N} \,.
\end{equation}
It is now convenient to split the coordinates into center-of-mass coordinates $y_a^\mu$, which are translation-invariant and commutative, and coordinates of the center of mass $x^\mu_\text{cm}$. We already found the algebra that these coordinates close,  Eq.~(\ref{XcmAlgebra}), and the main feature we would like to highlight is that the algebra realizes an action of the $x^\mu_\text{cm}$ generators on the $y_a^\mu$ ones, because the $x^\mu_\text{cm}$ generators close a subalgebra, and their commutators with $y_b^\nu$ gives a linear combination of $y_b^\nu$.
Consider now this fact:
\begin{equation}
e^{i k^a_\mu  (y^\mu_a + x^\mu_\text{cm} )}
= e^{i q^a_\mu  y^\mu_1 }  e^{i k^a_\mu x^\mu_\text{cm} }
\end{equation}
where $q^a_\mu  = q^a_\mu (k^a_0,k^a_1,k^a_2,k^a_3)$ is a certain function of $k^a_\mu$. The above equation is always true, and is a consequence of the Baker--Campbell--Hausdorff formula when $k^a_\mu x^\mu_\text{cm}$ belongs to the subalgebra acting on $k^a_\mu  y^\mu_a$ . Now consider this other identity, which is always true for Lie groups:
\begin{equation}
e^{i k^a_\mu x^\mu_\text{cm} } e^{i q^b_\mu  y^\mu_b }
= 
e^{i (k^a \triangleright q^b)_\mu  y^\mu_b } e^{i k^a_\mu x^\mu_\text{cm} } 
\end{equation}
where $\triangleright$ is the adjoint action of the group on itself. Finally, the subgroup properties imply the existence of an associative deformed sum of momenta $\boxplus : \mathbbm{R}^4 \times \mathbbm{R}^4 \to \mathbbm{R}^4$ which realizes the product of the subgroup generated by $x^\mu_\text{cm}$:
\begin{equation}
e^{i p_\mu  x^\mu_\text{cm} } e^{i q_\mu  x^\mu_\text{cm} }
=
e^{i (p \boxplus q)_\mu  x^\mu_\text{cm} } \,.
\end{equation}
Armed with the three identities listed above, we can rewrite Eq.~(\ref{NCFourier}) in the following form:
\begin{equation}
\begin{aligned}
f(x^\mu_a) &=  \int d^4 k^1 \dots d^4 k^N \,  \tilde f(k_\mu^a) \,
e^{i q^1_\mu  y^\mu_1}  e^{i k^1_\mu  x^\mu_\text{cm} }
e^{i q^2_\mu  y^\mu_2}  e^{i k^2_\mu  x^\mu_\text{cm} }
\dots
e^{i q^N_\mu  y^\mu_N}  e^{i k^N_\mu  x^\mu_\text{cm} } 
\\
&=  \int d^4 k^1 \dots d^4 k^N \,  \tilde f(k_\mu^a) \,
e^{i q^1_\mu  y^\mu_1}  
e^{i (k^1 \triangleright q^2)_\mu  y^\mu_2} 
\dots
e^{i (k^1 \boxplus k^2 \boxplus \dots \boxplus k^{N-1}) \triangleright q^N_\mu  y^\mu_N} 
 e^{i (k^1 \boxplus k^2   \boxplus \dots \boxplus k^N)_\mu  x^\mu_\text{cm} } \,.
\end{aligned}
\end{equation}
This proves that, if $\tilde f(k_\mu^a)  \propto \delta^{(4)}(k^1 \boxplus k^2   \boxplus \dots\boxplus k^N)$, the dependence on $x^\mu_\text{cm}$ completely disappears. A necessary condition for $f(x^\mu_a)$ to be $\kappa$-Poincar\'e-invariant is that $k^1 \boxplus k^2   \boxplus \dots\boxplus k^N =0$ so that the dependence on $x^\mu_\text{cm}$ drops. In fact, transforming all coordinates according to the coaction~(\ref{kappaCoaction}) we get
\begin{equation}
\begin{aligned}
f({x'}^\mu_a) =    \int d^4 k^1 \dots d^4 k^N \,  \tilde f(k_\mu^a) \,
e^{i q^1_\mu  \Lambda^\mu{}_\nu y^\nu_1}  
e^{i (k^1 \triangleright q^2)_\mu \Lambda^\mu{}_\nu y^\nu_2} 
\dots
e^{i (k^1 \boxplus k^2 \boxplus \dots \boxplus k^{N-1}) \triangleright q^N_\mu  \Lambda^\mu{}_\nu y^\nu_N}
\\ 
\cdot  e^{i (k^1 \boxplus k^2   \boxplus \dots \boxplus k^N)_\mu  \Lambda^\mu{}_\nu x^\nu_\text{cm} }
  e^{i (k^1 \boxplus k^2   \boxplus \dots \boxplus k^N)_\mu  a^\mu_\text{cm} } \,,
\end{aligned}
\end{equation}
and the $a^\mu$-dependent exponential disappears only if $k^1 \boxplus k^2   \boxplus \dots\boxplus k^N =0$. Therefore translation invariance alone ensures that $N$-point functions are commutative, because they are elements of the Abelian subalgebra generated by $y^\mu_a$.

Let us now take a deep dive into the structures that are necessary to build a consistent QFT on $\kappa$-Minkowski. We will begin with the properties of plane waves, which, as we already remarked, are Lie group elements, and can be mapped into points on a pseudo-Riemannian manifold, \emph{momentum space}. We will study all the structures that spacetime noncommutativity induces on said momentum space, and their relation. We will focus in particular on the issue of ordering and coordinate systems on momentum space: each ordering prescription of polynomials of noncommutative coordinates correspond to a choice of coordinates on momentum space, and changes of ordering coincide with diffemorphisms on momentum space. One of the guiding principles of our analysis will be that all physical quantities (and, in particular, $N$-point functions), will have to be independent of the ordering choice, and therefore they will have to be Riemannian invariants on momentum space.

From now on, we will focus on 1+1 spacetime dimensions, which simplify significantly the calculations, although everything we say can be  generalized to arbitrary dimensions.

\subsection{Plane waves paraphernalia}

In the $1+1$-dimensional case, it is convenient to rewrite the algebra~(\ref{Braided-kappa-Minkowski}) in lightcone coordinates:
\begin{equation}
x_a^\pm = x_a^0 \pm x_a^1 \,, \qquad x^0_a = \frac{x_a^+ + x_a^-}{2} \,, ~~ x^1_a = \frac{x_a^+ - x_a^-}{2} \,,
\end{equation}
the commutation relations take now the form
\begin{equation}\label{LightConeCoordinatesBraidedKappa}
[x^+_a , x^+_b ] = 2 i  (x^+_a - x^+_b) \,,  ~~ 
[x^+_a , x^-_b ] = 2 i x^-_b \,,  ~~ 
[x^-_a , x^+_b ] = - 2 i x^-_a \,,  ~~ 
[x^-_a , x^-_b ] = 0 \,.
\end{equation}
When $a=b$, the coordinates of a single point close the algebra
\begin{equation}\label{LightConeCoordinatesBraidedKappa_onepoint}
[x^+_a , x^-_a ] = 2 i x^-_a \,,
\end{equation}
which is identical (up to a normalization factor) to the timelike 1+1-dimensional $\kappa$-Minkowski algebra $\mathfrak{an}(1)$. The natural ordering prescriptions for polynomials then involve putting the $x^+_a$ coordinate to the right (resp. left) of $x^-_a$:
\begin{equation}
: (x^+_a)^n (x^-_a)^m :_\text{R}  =  (x^-_a)^m (x^+_a)^n \,,
\end{equation}
\begin{equation}
: (x^+_a)^n (x^-_a)^m :_\text{L}  =   (x^+_a)^n (x^-_a)^m \,,
\end{equation}
or symmetrizing them:
\begin{equation}
: (x^+_a)^n (x^-_a)^m :_\text{S}  =   \frac 1 2  (x^+_a)^n (x^-_a)^m + \frac 1 2 (x^-_a)^m (x^+_a)^n \,,
\end{equation}
or, also, Weyl-ordering them:
\begin{equation}
: (x^+_a)^n (x^-_a)^m :_\text{W}  =   \frac{1}{2^m} \sum_{r=0}^m (x^+_a)^{m-r} (x^-_a)^r \,.
\end{equation}
The linear maps $: \underline{~}  :_\text{R}$, $: \underline{~}  :_\text{L}$, $: \underline{~}  :_\text{S}$ and $: \underline{~}  :_\text{W}$ are \emph{Weyl maps}, that go from the algebra of commutative polynomials to the algebra~(\ref{LightConeCoordinatesBraidedKappa}), see, \emph{e.g.}~\cite{Agostini:2002de}. These maps are isomorphism from our noncommutative algebra of functions to the commutative one, and the commutation relations~(\ref{LightConeCoordinatesBraidedKappa_onepoint}) allow to translate from one map to the other, \emph{e.g.}:
\begin{equation}
: x^+_a x^-_a  :_\text{R} = : \left( x^+_a x^-_a + 2 i x^-_a  \right) :_\text{L} 
= : \left( x^+_a x^-_a +  i x^-_a  \right) :_\text{W} \,, 
\end{equation}
The linear nature of Fourier theory allows us to use these Weyl maps to map commutative Fourier-transfor{\-}mable functions (understood as functions on momentum space) to noncommutative functions with a certain ordering. As we will show, the same noncommutative function will admit different Fourier transforms, one for each choice of ordering, and these momentum space functions are related to each other by general coordinate transformations, \emph{i.e.}, diffeomorphisms of momentum space~\cite{Mercati:2018hlc}.

\subsubsection{Plane waves of a single coordinate}

For illustrative purposes, from now on we will work with right-ordered and Weyl-ordered functions, showing at each step how to translate  one description into the other. Again, keep in mind that our guiding principle is that no physical quantity should depend on the ordering choice.
Introduce the right-ordered plane waves, which provide a basis for Fourier theory:
\begin{equation}
\E_a[k] = e^{i k_- x_a^-} e^{i k_+ x_a^+} \,,
\end{equation}
they are labeled by $(k_-,k_+) \in \mathbbm{R}^2$, and are closed under Hermitian conjugation:
\begin{equation}
\E_a^\dagger [k] = \E_a[S(k)]  \,, \qquad S(k) = (- e^{2 k_+} k_-, - k_+) \,.
\end{equation}
The map $S: \mathbbm{R}^2 \to \mathbbm{R}^2$ is an involution ($S \circ S = \text{id}$), called \emph{antipode}. Since, as we remarked earlier, $\E_a[k]$ span the whole group $AN(1)$ associated to the Lie algebra $[x^+_a , x^-_a ] = 2 i x^-_a$, the map $S$ realizes the group inverse, and its properties follow from it. Another group axiom that can be represented as a map on the coordinates $k_\pm$ is the product:
\begin{equation}
\E_a [k] \E_a [q] = \E_a[k \oplus q ]  \,, \qquad k \oplus q  = (k_- + e^{-2 k_+} q_-, k_+ + q_+) \,,
\end{equation}
now the map $\oplus : \mathbbm{R}^2 \times \mathbbm{R}^2 \to \mathbbm{R}^2$ will be referred to as \emph{coproduct}, or \emph{momentum composition law}. Its properties follow from the axioms of Lie groups:
\begin{equation}
(k \oplus q ) \oplus p =  k \oplus ( q  \oplus p)  =  k \oplus  q  \oplus p \,, \qquad k \oplus S(k) = S(k) \oplus k = o \,, \qquad S(k \oplus q) = S(q) \oplus S(k) \,.
\end{equation}
The first rule expresses the associativity of $\oplus$, the second is the fact that $S$ is a bilateral inverse for $\oplus$, where $o=(0,0)$ are the coordinates of the origin of momentum space, and the third expresses the antihomomorphism property of the group inverse. 
The momentum-space origin $o$ is the neutral element for the composition law/coproduct:
\begin{equation}
o \oplus q  = q \oplus o = q \,, 
\end{equation}
and the plane waves with momentum $o$ are the identity element of the algebra:
\begin{equation}
\E_a[o] = 1 \,.
\end{equation}

\subsubsection{Translation-invariant products of two-point plane waves}

The product of two plane waves of different coordinates $\E_1[k]$, $\E_2[q]$ lies within the Abelian subalgebra of the functions of coordinate differences, $x^\mu_1 - x^\mu_2$ (\emph{i.e.}, it is translation-invariant), if the momenta of the two waves are the antipode of each other. There are four ways to combine two such waves:
\begin{equation}
\E_1 [k] \E_2^\dagger [k] \,, ~~
\E_1^\dagger [k] \E_2 [k] \,, ~~
\E_2 [k] \E_1^\dagger [k] \,, ~~
\E_2^\dagger [k] \E_1 [k] \,,
\end{equation}
the first and the third expressions are Hermitian conjugates, as are the second and fourth. Let us calculate explicitly now the functional form of the product of two waves:
\begin{equation}
\E_1 [k] \E_2 [q] =  e^{i k_- x_1^-} e^{i k_+ x_1^+} e^{i q_- x_2^-} e^{i q_+ x_2^+} \,,
\end{equation}
we want to order the expression by having all $x_a^-$ coordinates to the left, and all $x_a^+$ to the right. We need to commute $e^{i k_+ x_1^+}$ and $e^{i q_- x_2^-}$, where the coordinates $x_1^+$, $x_2^-$ close the subalgebra $[x_1^+,x_2^-]= 2 i x_2^-$ of the algebra~(\ref{LightConeCoordinatesBraidedKappa}), so, using the well-known $\kappa$-Minkowski commutation rule between exponentials~\cite{Mercati:2018hlc} (see Appendix~\ref{AppendixKappaAlgebra} Eq.~(\ref{RightOrderedCombinationLaw}) with $\kappa \to 1/2$):
\begin{equation}
e^{i k_+ x_1^+} e^{i q_- x_2^-} = e^{i e^{-2 k_+} q_- x_2^-} e^{i k_+ x_1^+} \,.
\end{equation}
The coordinates $x_a^-$ commute with each other~(\ref{LightConeCoordinatesBraidedKappa}), therefore our expression takes the form
\begin{equation}
\E_1 [k] \E_2 [q] =  e^{i \left( k_- x_1^-+ e^{-2 k_+} q_- x_2^- \right)} e^{i k_+ x_1^+} e^{i q_+ x_2^+} \,.
\end{equation} 
Since the coordinates $x_a^+$ close the subalgebra $[x_1^+,x_2^+]= 2 i (x_1^+ - x_2^+)$~(\ref{LightConeCoordinatesBraidedKappa}), it is convenient to make the linear redefinition
\begin{equation}
X = \frac{x_1^+ - x_2^+}{4} \,,
\qquad 
T = - \frac{x_1^+ + x_2^+}{4}
\,,
\qquad 
x^+_1 =  2(X-T) \,,
\qquad
x^+_2 =  -2(X+T) \,,
\end{equation}
so that we have another copy of the timelike $\kappa$-Minkowski algebra $[T , X] =  i X $, and use the rule to combine two Weyl-ordered $\kappa$-Minkowski waves~(\ref{WeylOrderedCombinationLaw}) (recall that we are using the convention $\kappa = 1$):
\begin{equation}
e^{ i \left( \alpha  T + \beta X \right)} e^{ i \left( \gamma  T + \delta X \right)} 
=
 e^{ i (\alpha+\gamma)  T + i \left(\frac{(\alpha+\gamma)}{1 - e^{- (\alpha+\gamma)}}\right) \left[ \left(\frac{1 - e^{- \alpha}}{\alpha}\right)  \beta + e^{-\alpha} \left(\frac{1 - e^{- \gamma}}{\gamma }\right)  \delta \right] X } 
 \,,
\end{equation}
since $e^{i k_+ x_1^+} e^{i q_+ x_2^+} = e^{i 2 k_+(X-T)} e^{- i 2 q_+ (X+T)}$\, we can obtain our desired expression by making the replacements $\alpha = - 2 k_+$, $\beta =  2 k_+$, $\gamma = - 2 q_+$, $\delta = - 2 q_+$. The result is:
\begin{equation}
\begin{aligned}
e^{i k_+ x_1^+} e^{i q_+ x_2^+}
&=
 e^{ -2 i ( k_+ + q_+)  T + 2i \left(\frac{ k_+ + q_+}{1 - e^{ 2 (k_+ + q_+)}}\right) \left[ \left(1 - e^{2 k_+}\right)   
 - e^{ 2 k_+} \left(1 - e^{ 2 q_+} \right)  \right] X } 
\\
&= 
e^{ 2 i ( k_+ + q_+)  \left[ \left( \frac{1 - 2 e^{2 k_+} + e^{ 2 (k_+ + q_+)} }{1 - e^{ 2 (k_+ + q_+)}} \right) X - T  \right] } 
 \,.
\end{aligned}
\end{equation}
Replacing the expressions for $X$ and $T$ we obtain the final expression:
\begin{equation}\label{ProductPlaneWaves}
\E_1 [k] \E_2 [q]   
=
e^{i \left( k_- x_1^-+ e^{-2 k_+} q_- x_2^- \right)}  e^{  i   \frac{k_+ + q_+}{1 - e^{ 2 (k_+ + q_+)}}  \left[ \left( 1 -  e^{2 k_+} \right) x^+_1 +  e^{2 k_+}  \left( 1- e^{ 2 q_+} \right) x^+_2 \right] } 
\,,
\end{equation} 
For $q = S(k) = (- e^{2 k_+} k_-, - k_+)$, the whole expression turns into:
\begin{equation}\label{TranslationInvariantPlaneWave1}
\E_1 [k] \E_2^\dagger [k] =   e^{i \xi_- \left( x_1^- - x_2^- \right)} 
e^{  i  \xi_+ \left( x^+_1 -  x^+_2 \right) } 
\,,
\end{equation} 
where we introduce the notation (which will be useful later):
\begin{equation}\label{LinearCoordinates}
\xi_- = k_- \,, \qquad \xi_+ = \left( \frac{e^{2 k_+} - 1}{2} \right) \,.
\end{equation}
The expression above depends only on $(x^\mu_1- x^\mu_2)$, as anticipated. Had we chosen to put the dagger on $\E_1[k]$, we would have obtained:
\begin{equation}\label{TranslationInvariantPlaneWave2}
\E_1^\dagger [k] \E_2 [k] =  
e^{ i \chi_- \left( x_1^- - x_2^- \right)} 
e^{  i   \chi_+  \left( x^+_1 - x^+_2 \right) } 
 \,,
\end{equation} 
where
\begin{equation}
\chi_- = - e^{2 k_+} k_- = S(\xi_-) \,, \qquad \chi_+ =  \left(\frac{ e^{-2 k_+} - 1}{2}\right) = S(\xi_+) \,.
\end{equation}
Notice how the functions $\xi_\pm(k_\pm)$ map $\mathbbm{R}^2$ into a half-plane of $\mathbbm{R}^2$, because $\xi_+ > - \frac 1 2$. This has significant consequences, which we will comment upon below. Here we just observe that, if one multiplies the two plane waves $\E_1[k]$ and $\E_2^\dagger[k]$, which have arbitrary frequencies $k_\mu \in \mathbbm{R}^2$, the resulting translation-invariant wave~(\ref{TranslationInvariantPlaneWave2}) cannot have any frequency. The coordinate differences $x_1^\mu - x_2^\mu$ in~(\ref{TranslationInvariantPlaneWave2}) are multiplied by frequencies that belong to a sub-region of $\mathbbm{R}^2$. 

The other two possible translation-invariant products of plane waves can be obtained from scratch with an analogous calculation, or, equivalently, by taking the Hermitian conjugate of the expressions (\ref{TranslationInvariantPlaneWave1}) and (\ref{TranslationInvariantPlaneWave2}), which simply amount to changing the sign of the exponents (or swapping coordinates 1 and 2), because the coordinate differences $(x^\mu_1- x^\mu_2)$  commute with each other:
\begin{equation}\label{TranslationInvariantPlaneWave3}
\E_2 [k] \E_1^\dagger [k]  = \left( \E_1 [k] \E_2^\dagger [k] \right)^\dagger  = 
e^{-i \xi_- \left( x_1^- - x_2^- \right)} 
e^{ - i \xi_+ \left( x^+_1 -  x^+_2 \right) } 
 \,,
\end{equation} 
\begin{equation}\label{TranslationInvariantPlaneWave4}
\E_2^\dagger [k] \E_1 [k] = \left( \E_1^\dagger [k] \E_2 [k] \right)^\dagger = 
e^{ - i \chi_-  \left( x_1^- - x_2^- \right)} 
e^{ - i   \chi_+ \left( x^+_1 - x^+_2 \right) } 
 \,.
\end{equation}

\subsubsection{Plane waves in different bases/orderings}

Define now the Weyl-ordered plane waves as
\begin{equation}
\F_a[q_- , q_+] = e^{i q_- x^-_a + i q_+ x^+_a} 
=
\E_a \left[ \textstyle  \left(\frac{1 - e^{-2 q_+}}{2 q_+} \right) q_- , q_+\right] 
\,,
\end{equation}
where the last expression comes from Eq.~(\ref{WeylOrderedCombinationLaw0}) with $\kappa \to 1/2$.
If $k_\pm$ are the frequencies of a right-ordered plane wave $\E(k_\pm)$, and $q_\pm$ are those of a Weyl-ordered wave $\F(q_\pm)$, their relation is:
\begin{equation}\label{Right-vs-Weyl-coordinates}
k_- =  \left( \frac{1 -  e^{-2 q_+}}{2 q_+} \right) q_- \qquad k_+ =  q_+  \,, \qquad
q_- = \left( \frac{2 k_+}{1 -  e^{-2 k_+}} \right) k_- \qquad q_+ =  k_+
\,,
\end{equation}
which implies that $\E(k_\pm)=\F(q_\pm)$.

Let us now look at the composition law of Weyl-ordered waves:
\begin{equation}
\F_1[k]  \F_1[q] = \F_1[ k \oplus' q ]
 \,,
\end{equation}
the map $\oplus'$ is explicitly calculated in Appendix~\ref{AppendixKappaAlgebra}: replacing $\kappa$ with $1/2$ in Eq.(\ref{WeylOrderedCombinationLaw}) we get
\begin{equation}
\begin{aligned}
(k \oplus' q)_- &=  \left(\frac{2(k_++q_+)}{1 - e^{- 2(k_++q_+)}}\right) \left[ \left(\frac{1 - e^{-2 k_+}}{2k_+}\right)  k_- + e^{- 2 k_+ } \left(\frac{1 - e^{- 2  q_+ }}{2 q_+ }\right)  q_- \right] \,,
\\
(k \oplus' q)_+ &= k_+ + q_+
\,.
\end{aligned}
\end{equation}
The antipode map:
\begin{equation}
\F^\dagger[q] = \F[S'(q)] ~ \Rightarrow ~ S'[k] = - k \,,
\end{equation}
can also be calculated by combining the right-ordered antipode, $S(k) = (- e^{2 k_+} k_-, - k_+)$, with the coordinate change~(\ref{Right-vs-Weyl-coordinates}), which we will call $\phi$:
\begin{equation}
\begin{aligned}
S'(q) &=  (\phi^{-1} S \circ \phi )(q)  =   \phi^{-1} \left[  \left( \frac{2 S(k)_+}{1 -  e^{-2 S(k)_+}} \right) S(k)_- , S(k)_+  \right]
\\
&=  \left(  - \left( \frac{ 2 k_+}{1 -  e^{-2 k_+}} \right) k_- , - k_+  \right) \Big{|}_{\tiny \begin{array}{l}
k_- \to  \left( 1 -  e^{-2 q_+}\right)/2 q_+
\\
k_+ \to q_+ 
\end{array} } = (- q_- , - q_+ ) \,,
\end{aligned}
\end{equation}
which confirms the above result that the $S'$ map just puts a minus in front of both components of $q$. Finally, we can check where the origin of momentum space is mapped by the coordinate change: $o'(q) = \phi^{-1} \circ o \phi )(q) = (0,0)$, which is consistent with the fact that $\F_a[q] \to 1$  when $q_- = q_+ = 0$.

We are now ready to study the products of plane waves of different points. The calculation is similar to that for right-ordered waves:
\begin{equation}
\begin{aligned}
\F_1[k] \F_2 [q] &=  e^{i k_- x^-_1 + i k_+ x^+_1}  e^{i q_- x^-_2 + i q_+ x^+_2} 
= e^{ i  \left( \frac{1-e^{-2k_+}}{2k_+} \right) k_-  x^-_1 }e^{i k_+ x^+_1}  e^{i \left( \frac{1-e^{-2q_+}}{2 q_+} \right)q_- x^-_2 }e^{i q_+ x^+_2}
\\
&= e^{ i  \left( \frac{1-e^{-2 k_+}}{2 k_+} \right) k_-  x^-_1 }e^{i e^{-2 k_+} \left( \frac{1-e^{-2 q_+}}{2 q_+} \right)q_- x^-_2 } e^{i k_+ x^+_1}  e^{i q_+ x^+_2}
\\
&= e^{ i   \left( \frac{1-e^{-2 k_+}}{2 k_+} \right) k_-  x^-_1 + i e^{-2 k_+} \left( \frac{1-e^{-2 q_+}}{2 q_+} \right)q_- x^-_2 } e^{  i   \frac{k_+ + q_+}{1 - e^{ 2 (k_+ + q_+)}}  \left[ \left( 1 -  e^{2 k_+} \right) x^+_1 +  e^{2 k_+}  \left( 1- e^{ 2 q_+} \right) x^+_2 \right] } \,,
\end{aligned}
\end{equation}
and setting $q = S'(k)$ gives a translation-invariant product of waves:
\begin{equation}
 \F_1[k] \F_2^\dagger [k] =  e^{ i   \left( \frac{1-e^{-2 k_+}}{2 k_+} \right) k_- ( x^-_1  - x^-_2) + \frac{i}{2} \left(e^{2 k_+}-1\right)( x_1^+ - x_2^+ )} \,.
\end{equation}


\subsubsection{Lorentz transformations of momenta and plane waves}
\label{LorentzTransformationsOfMomentaSec}

Under the Poincar\'e coaction~(\ref{kappaCoaction}), our plane waves transform in the following way:
\begin{equation}\label{FormOfLorentTransformsOfPlaneWave}
\E'_a[k] = \E_a[\lambda(k,\Lambda)] \A[k] \,,
\end{equation}
where $\lambda$ is a nonlinear representation of the Lorentz group:
\begin{equation}
\lambda : \mathbbm{R}^2 \times SO(1,1) \to \mathbbm{R}^2 \,, 
~~~  \,,
~~
\lambda(\lambda(k,\Lambda),\Lambda') = \lambda(k,\Lambda^\mu{}_\rho \Lambda'^\rho{}_\nu) \,,
~~
\Lambda(k , \delta^\mu{}_\nu) = k \,,
~~
\lambda(o,\Lambda)=0 \,,
\end{equation}
and
\begin{equation}
\A[k] = e^{i k_- a^- } e^{i k_+ a^+} \,,
\end{equation}
is an ordered plane wave of the translation parameters (the notation $a^\pm = a^0 \pm a^1$ should be clear at this point).

In order to calculate $\lambda (k,\Lambda)$, we could exploit the homomorphism property of the coaction and Poincar\'e-transform the two sides of Eq.~(\ref{TranslationInvariantPlaneWave1}), which depends only on the difference between coordinates and therefore is translation-invariant:
\begin{equation}  \label{eprimeedagprime}
\E_1' [k] \E_2'^\dagger [k] =  e^{i \xi_- \left(  x_1'^- - x_2'^- \right)} e^{i \xi_+ \left(x_1'^+-x_2'^+\right) } =
 e^{i \lambda(k,\Lambda)_- \left(  x_1^- - x_2^- \right)} e^{i \left(\frac{e^{2 \lambda(k,\Lambda)_+}-1}{2}\right) \left(x_1^+-x_2^+\right) } \,,
\end{equation}
since
\begin{equation}
\left(  x_1'^- - x_2'^- \right) = e^{-\omega} \left(  x_1^- - x_2^- \right) \,,
\qquad
\left(  x_1'^+ - x_2'^+ \right) = e^{+\omega} \left(  x_1^+ - x_2^+ \right) \,,
\end{equation}
where $\omega$ is the rapidity, $\Lambda^0{}_0=\Lambda^1{}_1= \cosh \omega$,  $\Lambda^0{}_1=\Lambda^1{}_0= \sinh \omega$. Consistency demands that
\begin{equation}
\lambda(k,\xi)_- = e^{-\omega} \xi_- \,, \qquad
\frac{e^{2 \lambda(k,\xi)_+}-1}{2} = e^{+\omega} \xi_+ \,,
\end{equation}
which admits the solution
\begin{equation}\label{LorentzTransformMomentum}
\lambda(k,\xi)_- = e^{-\omega} \xi_-   = e^{-\omega} k_- \,, \qquad
\lambda(k,\xi)_+ = \frac 1 2 \log \left( 1+ 2 e^{\omega} \xi_+ \right)  = \frac 1 2 \log \left[ 1+  e^{\omega}  \left( e^{2 k_+} - 1\right) \right] \,.
\end{equation}
The `$-$' component transforms in an undeformed way, while the transformation of the `$+$' component is nonlinear. In a power series in the momentum:
\begin{equation}
\lambda(k,\xi)_+ = k_+ e^{\xi }- k_+^2 e^{\xi } \left(e^{\xi }-1\right)+ \mathcal{O}(k_+^3) \,.
\end{equation}
One can easily verify that $\lambda(k,\xi)$ is a representation of the Lorentz group:
\begin{equation}
\lambda(k,\xi + \xi') = \lambda(\lambda(k,\xi),\xi') \,, \qquad  \lambda(k,0) = k  \,, 
\end{equation}
and that it leaves the origin unchanged: $\lambda(o,\xi) =0$.

We could be content with having found the map $\lambda$ from Eq.~(\ref{TranslationInvariantPlaneWave1}), but we used Eq.~(\ref{FormOfLorentTransformsOfPlaneWave}) dictating the form of the Poincar\'e transformation of a plane wave, and therefore we do not yet have a proof. We need to go through the pain of proving it with a direct calculation, which however will allow us to derive the form of $\lambda(k,\Lambda)$ directly, providing a check that the Poincar\'e coaction is indeed a homomorphism ot the coordinate algebra, showing that the two sides of Eq.~(\ref{TranslationInvariantPlaneWave1}) transform in the same way.

First of all, we need to write the lightlike $\kappa$-Poincar\'e commutation relations~(\ref{kappaPoincareGroup}) in a more convenient way, adapted to the $1+1$-dimensional lightlike case:
\begin{equation}\label{1+1D-lightlik-kappa-Poincare}
[a^+ , \omega ] = 2 i \left( e^\omega -1 \right) \,, \qquad [a^- ,\omega] = 0 \,,
\end{equation}
where $a^\pm = a^0 \pm a^1$, and $\omega$, again, is the rapidity $\Lambda^0{}_0 = \cosh \omega$. We are first interested in the adjoint action of exponentials of the translation parameters $\A[k]= e^{i k_- a^-} e^{i k_+ a^+}$ on arbitrary functions of the rapidity. Since $a^-$ commutes with $\omega$ it will go though, instead $a^+$ is canonically conjugate to the coordinate $\rho = \frac 1 2 \log \left( e^{-\omega} -1 \right)$:
\begin{equation}
[  a^+ , \frac 1 2 \log \left( e^{-\omega} -1 \right) ] = [a^+,\rho]=i \,,
\end{equation}
therefore it acts like a translation for $\rho$:
\begin{equation}\label{DefinitionBackreaction}
e^{i k_+ a^+} f(\rho) e^{-i k_+ a^+} = f(\rho - k_+) \,,
\end{equation}
which corresponds to a nonlinear action on the coordinate $\omega$:
\begin{equation}\label{Backreaction}
e^{i k_+ a^+} f(\omega) e^{-i k_+ a^+} = f\left[ -\log \left( 1 + ( e^{-\omega} -1 ) e^{-2 k_+}\right) \right] \,.
\end{equation}
This has been sometimes described~\cite{Majid:1994cy,Gubitosi:2013rna} as a `backreaction'  of the momenta on the Lorentz sector, part of the `bicrossproduct' structure of the $\kappa$-Poincar\'e group, represented as a right action:
\begin{equation}
\triangleleft : SO(1,1) \times \mathbbm{R}^2 \to SO(1,1) \,, \qquad \omega \triangleleft k =  -\log \left( 1 + ( e^{-\omega} -1 ) e^{-2 k_+} \right) \,.
\end{equation}
From the definition of $\triangleleft$~(\ref{DefinitionBackreaction}), it follows that it is a coalgebra homomorphism for the coproduct, \emph{i.e.}
\begin{equation}
(f(\Lambda) \triangleleft p) \triangleleft q =f(\Lambda) \triangleleft ( p \oplus q ) \,, \qquad f(\Lambda) \triangleleft o = f(\Lambda) \,,
\end{equation}
and the adjoint action of translations on rapidities can be written:
\begin{equation}\label{AdjointBackreaction}
\A[k] f(\omega) \A^\dagger[k] = f(\omega \triangleleft k)\,, \qquad 
\A^\dagger[k] f(\omega) \A[k] = f(\omega \triangleleft S[k])\,,\,.
\end{equation}

The next step is to calculate the opposite action, the adjoint action of any function $f(\omega)$ on the translation parameters $a^\mu$. The $a^-$ parameter commutes with $\omega$ and will therefore be invariant. Regarding $a^+$, from the commutation relations with $\omega$, and its immediate consequence 
$[a^+,g(\omega)]=2i(e^\omega-1)g'(\omega)$,
we deduce
\begin{equation}
e^{f(\omega)} a^+ = \left( a^+ - 2 i \left( e^\omega -1 \right) f'(\omega) \right) e^{f(\omega)} \,,
\end{equation}
iterating the procedure:
\begin{equation}
\begin{aligned}
e^{f(\omega)} (a^+)^2 &= \left( a^+ - 2 i \left( e^\omega -1 \right) f'(\omega) \right) e^{f(\omega)} a^+  = \left( a^+ - 2 i \left( e^\omega -1 \right) f'(\omega) \right)^2 e^{f(\omega)} \,,
\\
e^{f(\omega)} (a^+)^3 &=  \left( a^+ - 2 i \left( e^\omega -1 \right) f'(\omega) \right)^3 e^{f(\omega)} \,,
\\
\vdots
\\
e^{f(\omega)} (a^+)^n &=  \left( a^+ - 2 i \left( e^\omega -1 \right) f'(\omega) \right)^n e^{f(\omega)} \,,
\end{aligned}
\end{equation}
 by induction, we get
 \begin{equation}\label{AdActionXiOnA}
 e^{f(\omega)} e^{i k_+ a^+} = e^{i k_+ \left( a^+ - 2 i \left( e^\omega -1 \right) f'(\omega) \right)} e^{f(\omega)} \,.
 \end{equation}

Now consider the plane wave $\E_1[k]$ and apply a $\kappa$-Poincar\'e transformation to it:
\begin{equation}\label{LorentzTransformPlaneWave1}
\E_1'[k] = e^{i k_- ( e^{-\omega} x^- + a^-)}e^{i k_+ ( e^{\omega} x^+ + a^+)} \,,
\end{equation}
using the commutativity of $a^-$ and $\omega$,
\begin{equation}\label{LorentzTransformPlaneWave2}
e^{i k_- ( e^{-\omega} x^- + a^-)}e^{i k_+ ( e^{\omega} x^+ + a^+)} = e^{i k_- e^{-\omega} x^-} e^{i k_- a^-} e^{i k_+ (e^{\omega} x^+ + a^+)} \,.
\end{equation}
Consider now Eq.~(\ref{AdActionXiOnA}), for the following choice of function: $f(\omega) =  \frac{i}{2}   \log \left(e^{\omega }-1\right) x^+$. It takes the form:
\begin{equation}
 e^{i k_+ \left( a^+ + e^\omega x^+ \right)}  =  e^{ \frac{i}{2}   \log \left(e^{\omega }-1\right) x^+} e^{i k_+ a^+} e^{-\frac{i}{2}   \log \left(e^{\omega }-1\right) x^+} \,,
\end{equation}
which can be immediately substituted in~(\ref{LorentzTransformPlaneWave2}):
\begin{equation}\label{LorentzTransformPlaneWave3}
\E_1'[k] =  e^{i k_- e^{-\omega} x^-} e^{i k_- a^-}  e^{ \frac{i}{2}   \log \left(e^{\omega }-1\right) x^+} e^{i k_+ a^+} e^{-\frac{i}{2}   \log \left(e^{\omega }-1\right) x^+} \,.
\end{equation}
Now we want to bring the exponential $e^{i k_+ a^+}$ to the right, with the help of Eq.~(\ref{Backreaction}):
\begin{equation}\label{LorentzTransformPlaneWave4}
\begin{gathered}
 e^{i k_- e^{-\omega} x^-} e^{i k_- a^-}  e^{ \frac{i}{2}   \log \left(e^{\omega }-1\right) x^+} e^{i k_+ a^+} e^{-\frac{i}{2}   \log \left(e^{\omega }-1\right) x^+} = 
 \\
  e^{i k_- e^{-\omega} x^-} e^{i k_- a^-}  e^{ \frac{i}{2}   \log \left(e^{\omega }-1\right) x^+}  e^{-\frac{i}{2}   \log \left(\frac{ (1 - e^{-\omega}) e^{-2 k_+} }{1 + ( e^{-\omega} -1 ) e^{-2 k_+}} \right) x^+} e^{i k_+ a^+}=
 \\
  e^{i k_- e^{-\omega} x^-}  e^{ \frac{i}{2}   \log \left(e^{\omega }-1\right) x^+}  e^{-\frac{i}{2}   \log \left(\frac{ (1 - e^{-\omega}) e^{-2 k_+} }{1 + ( e^{-\omega} -1 ) e^{-2 k_+}} \right) x^+} e^{i k_- a^-}  e^{i k_+ a^+}
   \\
  e^{i k_- e^{-\omega} x^-}    e^{-\frac{i}{2}   \log \left(\frac{ e^{-\omega } e^{-2 k_+} }{1 + ( e^{-\omega} -1 ) e^{-2 k_+}} \right) x^+} e^{i k_- a^-}  e^{i k_+ a^+}=
    \\
  e^{i k_- e^{-\omega} x^-}    e^{ \frac{i}{2}   \log \left[ 1 +  e^{\omega } ( e^{2 k_+} - 1) \right] x^+} e^{i k_- a^-}  e^{i k_+ a^+} = \E_1[\lambda(k,\omega)] \A[k] 
 \,,
\end{gathered}
\end{equation}
which reproduces the formula~(\ref{LorentzTransformMomentum}) for $\lambda(k,\omega)$.

Consider the transformation rule of the translation-invariant products of two waves~(\ref{TranslationInvariantPlaneWave1}), $\E_1 [k] \E_2^\dagger [k]$. As we have seen above, the coordinate differences $x^\mu_1 - x^\mu_2$ transform following an underformed, 
linear Lorentz transformation, ${x'}^\mu_1 - {x'}^\mu_2 = \Lambda^\mu{}_\nu (x^\mu_1 - x^\mu_2)$, 
and the functions $\xi_\mu(k)$ appearing in front of $x^\mu_1 - x^\mu_2$ transform according to the (inverse) undeformed Lorentz transformation $\xi_\mu'(k) = \Lambda^\nu{}_\mu \xi_\nu(k)$, which can also be written as a transformation of the momentum parameter $k^\mu$, but in this case the transformation is nonlinear. In other words, $\xi_\mu$ provides a linear representation of the Lorentz group:
\begin{equation}
\xi_\mu [\lambda(k,\xi)] =  \xi_\nu[k]  \Lambda^\nu{}_\mu  \,.
\end{equation}
If we now consider the transformation law of the other translation-invariant product of waves, that is not just the Hermitian conjugate of the first one, Eq.~(\ref{TranslationInvariantPlaneWave2}), we obtain
\begin{equation}
\begin{aligned}
{\E_1'}^\dagger [k] \E_2' [k] &=  \A^\dagger [k] \E_1^\dagger [\lambda(k,\omega)] \E_2 [\lambda(k,\omega)] \A[k] 
\\
&= \A^\dagger [k]  \E_1 [S[\lambda(k,\omega)]]  \E_2 [\lambda(k,\omega)] \A[k] .
\end{aligned} 
\end{equation} 
Notice the following important identity:
\begin{equation}
\lambda(k,\omega) = S[\lambda(S[k],\omega \triangleleft k))]
\end{equation}
which implies also $S[\lambda(k,\omega)] = \lambda(S[k],\omega \triangleleft k))$, which can be used in the expression above:
\begin{equation}
\begin{aligned}
\A^\dagger [k]  \E_1 [S[\lambda(k,\omega)]]  \E_2 [\lambda(k,\omega)] \A[k] 
&= \A^\dagger [k]  \E_1 [\lambda(S[k],\omega \triangleleft k))]  \E_2 [\lambda(k,\omega)] \A[k]
\\
&=  \E_1 [\lambda(S[k],\omega \triangleleft k \triangleleft S[k])] \A^\dagger [k]  \E_2 [\lambda(k,\omega)] \A[k]
\\
&=  \E_1 [\lambda(S[k],\omega)]  \E_2 [\lambda(k,\omega \triangleleft S[k])]  \A^\dagger [k] \A[k]
\\
&=  \E_1^\dagger [\lambda(k,\omega \triangleleft S[k])]  \E_2 [\lambda(k,\omega \triangleleft S[k])] \,.
\end{aligned} 
\end{equation} 
Notice that the transformation rule of $k_\mu$ is not $k_\mu \to \lambda_\mu(k,\omega \triangleleft k)$, as it in the case of $\E_1 [k] \E_2^\dagger [k]$. The trasformation rule is instead
\begin{equation}
k_\mu \to \lambda_\mu(k,\omega \triangleleft S[k])\,,
\end{equation}
the novelty being in the transformed rapidity which is now $\omega \triangleleft S[k]$.  A direct calculation confirms that this particular rule makes the function $\chi_\mu [k]$ transform linearly:
\begin{equation}
\chi_\mu [\lambda(k,\omega \triangleleft S[k])] =  \chi_\nu[k] \Lambda^\nu{}_\mu \,,
\end{equation}
so that
\begin{equation}\label{TransformationLawChiWaves}
{\E_1'}^\dagger [k] \E_2' [k]
=\
e^{i \chi_\mu [k] \Lambda^\mu{}_\nu (x^\nu_1 - x^\nu_2)} = 
e^{i \chi_\mu [\lambda(k,\omega \triangleleft k)] (x^\mu_1 - x^\mu_2)} = 
\E_1^\dagger [\lambda(k,\omega \triangleleft k)] \E_2 [\lambda(k,\omega \triangleleft k)] \,.
\end{equation}

\subsection{Geometry of momentum space}

In~\cite{Lizzi:2020tci} we studied the general theory of $\kappa$ momentum spaces. Following the techniques illustrated in that paper, one can study the geometries of momentum space that are compatible with the lightlike $\kappa$-Minkowski space we are interested in. For our purposes, however, it is sufficient to observe that the coordinates $\xi_\pm$ transform linearly (like light-cone coordinates) under momentum-space Lorentz transformations, and therefore said transformations will leave invariant the following light-cone-coordinates Minkowski metric:
\begin{equation}
ds^2 =  d \xi_- d \xi_+  \,.
\end{equation}
In right-ordered coordinates $k_\pm$, which are related to $\xi_\pm$ by Eq.~(\ref{LinearCoordinates}),  this metric reads
\begin{equation}
ds^2 =  e^{2 k_+} \, d k_- d k_+ \,.
\end{equation}
As we observed after showing Eq.~(\ref{LinearCoordinates}),  the functions $\xi_\pm$ do not represent a map from $\mathbbm{R}^2$ to  $\mathbbm{R}^2$. They rather map  $\mathbbm{R}^2$ to the semiplane  $\xi_+  > - \frac 1 2$. The border $\xi_+ = -1/2$ of our coordinate system coincides with a lightlike line.
\begin{figure}[htb]\center
\includegraphics[width=0.4\textwidth]{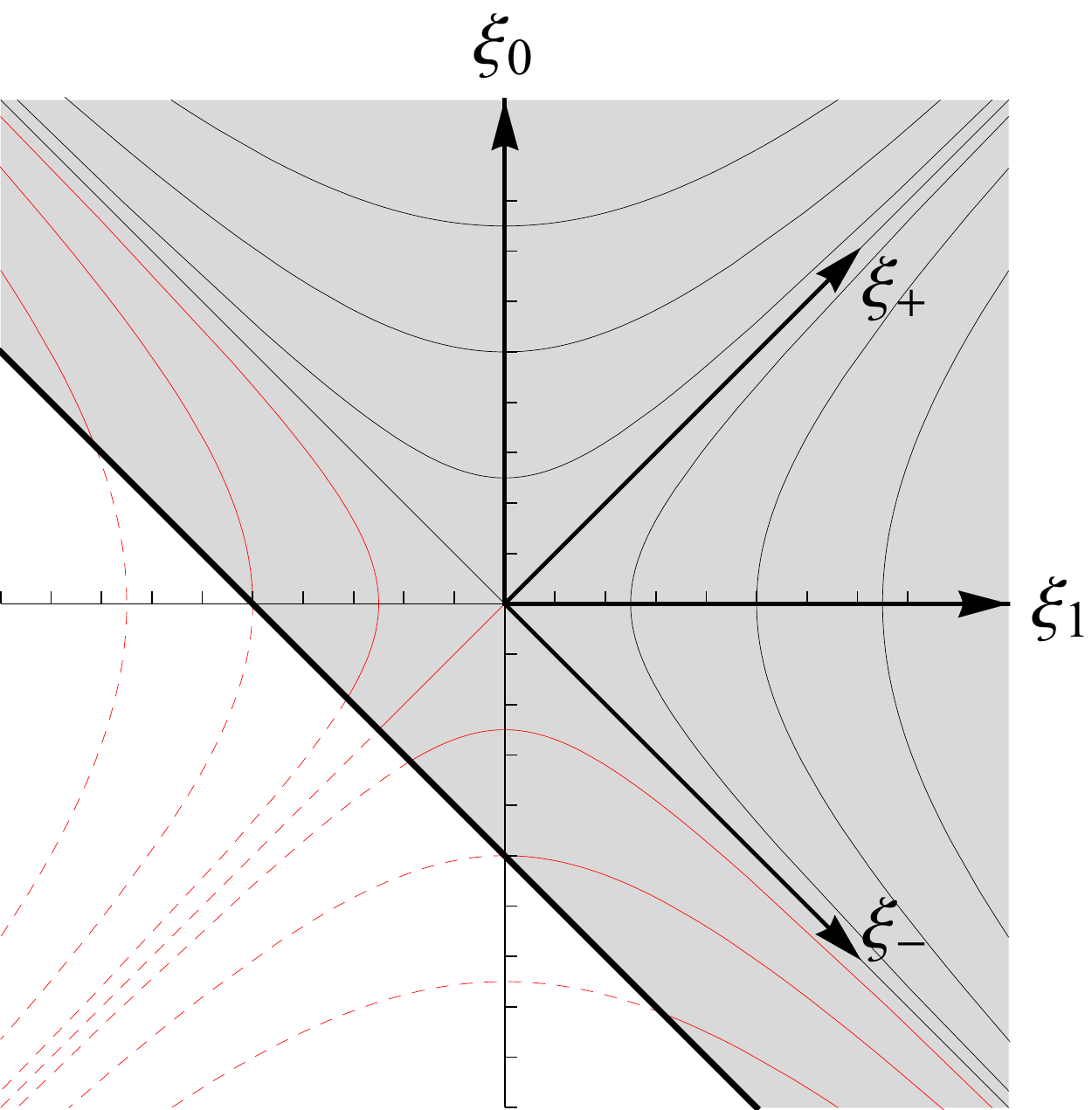}
\caption{\sl The momentum space of $1+1$-dimensional lightlike $\kappa$-Minkowski.}
\end{figure}

The presence of a finite border implies that our momentum space, despite being \emph{locally} Lorentz-invariant, is not so globally. This is reflected also in the form of the Lorentz trasnsformations of $k_\pm$, Eq.~(\ref{LorentzTransformMomentum}), which become singular at a finite value of $\xi$ when $e^{2 k_+} <1$, and the argument of the logarithm in $\lambda_+(k,\xi) = \frac 1 2 \log \left[ 1+ e^\xi \left(e^{2 k_+}-1\right) \right]$ is negative for all values of $\xi$ above $-\log \left( 1 - e^{2 k_+} \right)$. The situation is perfectly analogue to that of  `timelike' 
$\kappa$-Minkowski, with its half-de Sitter momentum space whose border can be reached with a finite Lorentz transformation. In that model, a way out of this Lorentz-breaking feature was to assume a different global topology for momentum space, by quotienting it by a reflection in the ambient space, thereby obtaining an \emph{elliptic} de Sitter momentum space, which is closed under Lorentz transformations. It's not obvious whether we can do something like that here.

It is easy to verify that the transformation $k_\pm \to \lambda(k,\xi)_\pm$ is an isometry of $ds^2$. 

The geodesics of momentum space are obviously straight lines in the coordinates $\xi_\pm$, which in coordinates $k_\pm$ are:
\begin{equation}
k_-(s) = \alpha \, s + k_-^0 \,, \qquad  k_+(s) = \frac 1 2 \log \left( \beta \, s + e^{2 k_+^0} \right) \,.
\end{equation}
The geodesic distance between the origin $o = (0,0)$ and the point $(k^1_-,k^1_+)$, along the geodesic connecting $o$ to $k^1_\mu$, $k_-(s) = k^1_- \, s$, $ k_+(s) = \frac 1 2 \log \left[ (e^{2 k_+^1} -1) \, s + 1 \right]$, is then:
\begin{equation}
\int_0^1 \sqrt{ 2  k^1_- (e^{2 k_+^1} -1)} ds = \sqrt{ 2  k^1_- (e^{2 k_+^1} -1)} \,.
\end{equation}
We can define a mass-shell operator as any function of the Geodesic distance (the difference between different choices of the function will amount to a nonlinear redefinition of the mass:
\begin{equation}\label{MassCasimir}
\mathcal{C} = k_- (e^{2 k_+} -1)  = \xi_- \xi_+ \,,
\end{equation}
it is easy to see how $\mathcal{C}$ is Lorentz-invariant.

\subsubsection{Mass shells}

The $k_\pm$ coordinates are deformations of light-cone coordinates. In special relativity, if we want to describe the mass shells through dispersion relations, light-cone coordinates are a bit different from the familiar energy-momentum ones. In terms of energy $E$ and momentum $p$, the mass-shell condition reads $E^2 - p^2 = m^2$, and solving this with respect to $E$ gives the two dispersion relations of positive- and negative-frequency waves: $E = \pm \sqrt{m^2 + p^2}$. In light-cone coordinates the mass shell is
\begin{equation} \label{SRmassShell}
p_+ p_- = m^2 \,,
\end{equation}
and solving with respect to one of the coordinates, \emph{e.g.} $p_-$, gives one single solution: $p_- = \frac{m^2}{2p_+}$. This is sufficient to describe both positive- and negative-frequency solutions with one single function: the former correspond to positive values of $p_+$, and the latter to $p_+<0$. In the case of imaginary mass, $m^2$ changes sign in  Eq.~(\ref{SRmassShell}), and the mass-shell we are describing are the tachionic ones.

Our $\kappa$-deformation does not change the basic qualitative picture: the mass-shell relation is (the normalization is chosen in order to match Eq.~(\ref{SRmassShell}) in the $\kappa \to 0 $ limit):
\begin{equation}\label{disprel1}
\mathcal{C}(k) = \frac 1 2 k_- \left(e^{2k_+} -1 \right) =  m^2 \,,
\end{equation}
which can be solved for $k_-$ as:
\begin{equation}
k_- = \omega_r(k_+) =  \frac{2 m^2}{e^{2k_+} -1 } \,,
\end{equation}
and the positive-frequency mass-shells correspond values of $k_+ > 0$, while the negative-frequency ones correspond to $k_+ <0$. Notice that $\omega_r \in (-\infty,-m^2) \cup (0 , \infty) $.

If we decided to use Weyl-ordered waves instead of right-ordered ones, the mass shell function would take a different form. Using the relation~(\ref{Right-vs-Weyl-coordinates}) between these two coordinate systems, we get the following form for the mass-shell function:
\begin{equation}\label{disprel2}
\mathcal{C}'(q)  = \frac 1 4  \left(e^{2q_+} -1 \right)^2 \frac{q_-}{q_+} =  m^2 \,,
\end{equation}
and the dispersion relation now takes the form
\begin{equation}
q_- = \omega_w(q_+) =  \frac{4 m^2 \, q_+ }{\left(e^{2q_+} -1 \right)^2} \,,
\end{equation}
again, $q_+<0$ describes positive-frequency waves, and $q_+<0$ negative-frequency ones.

\begin{figure}[htp]\center
\includegraphics[width=0.4\textwidth]{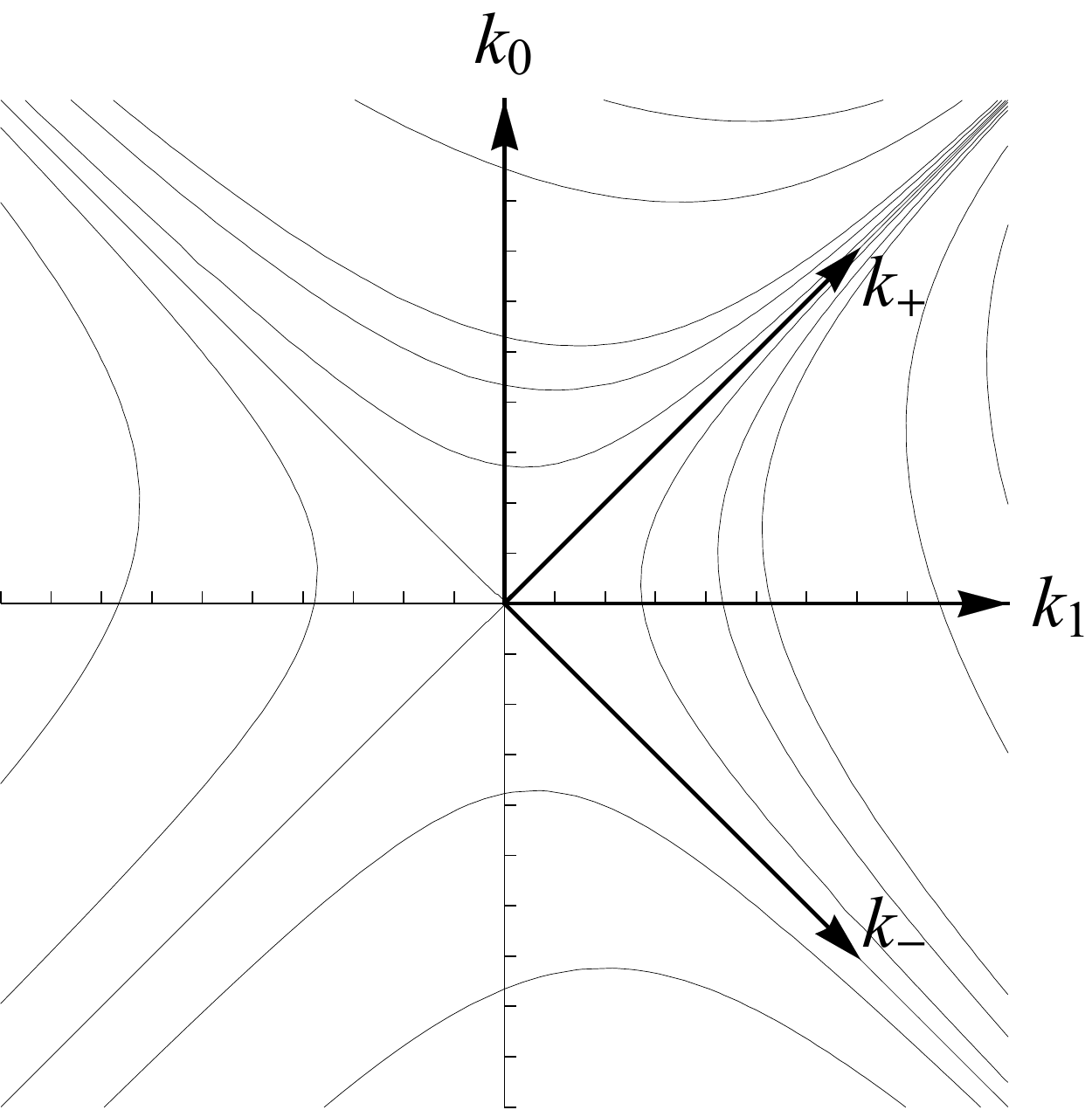}\qquad
\includegraphics[width=0.4\textwidth]{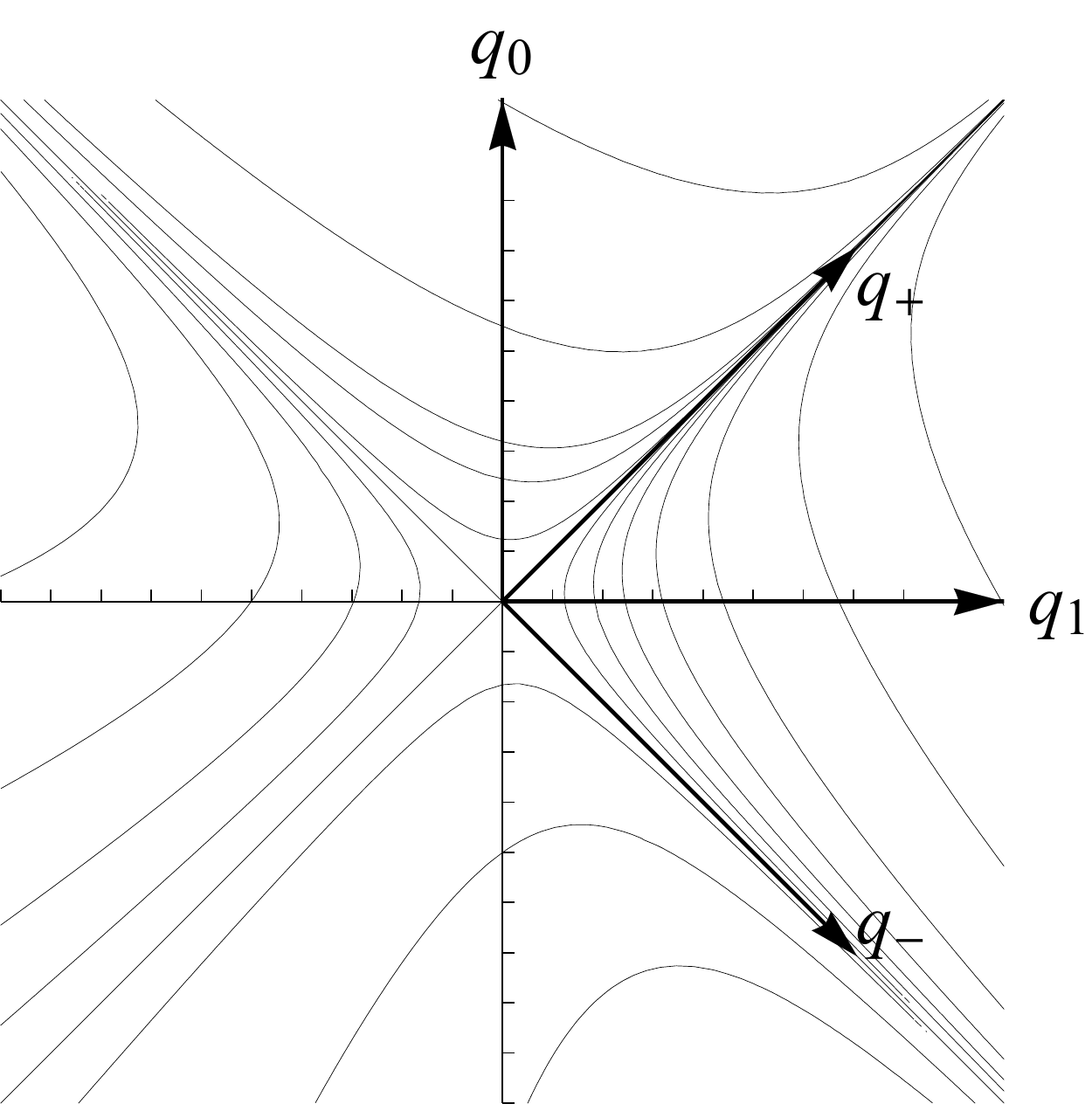}
\caption{\sl The dispersion relations~\eqref{disprel1} (left) and~\eqref{disprel2} (right).}
\end{figure}

\subsection{The $\kappa$-Klein--Gordon equation}
\label{Sec-k-KleinGordonEq}

Consider the following equation:
\begin{equation}\label{kappaKGEq}
\mathcal{C} \triangleright \phi(x_a) = m^2   \phi(x_a) \,,
\end{equation}
where the Casimir operator's action on noncommutative functions is defined in Fourier-transform as
\begin{equation}
\mathcal C \triangleright f(x_a) = \int d^2 k  \sqrt{-g(k)}\, \tilde f(k) \, \mathcal C (k)  \E_a[k] \,, \qquad 
f(x_a) = \int d^2 k  \sqrt{-g(k)}\, \tilde f(k) \, \E_a[k]
\,.
\end{equation}
The generic solution to Eq.~(\ref{kappaKGEq}) is
\begin{equation}
\begin{aligned}
  \phi (x_a) &= \int d^2 k \sqrt{-g(k)} \ \delta \left( \mathcal{C}(k) - m^2 \right)   \tilde \phi (k) \E_a[k] = \int d^2 k \sqrt{-g(k)} \frac{ \delta \left( k_- - \omega_r(k_+) \right) }{ \frac 1 2 \left| e^{2 k_+} - 1\right| } \, \tilde \phi (k) \E_a[k] \,, 
\end{aligned}
 \end{equation} 
 we can now split the function $ \tilde \phi (k)$  according to its values on the two mass-shells, the Lorentz-invariant one with $k_+ >0$ and the other one with $k_+<0$:
\begin{equation}
 \tilde \phi (k) = a(k_+) \Theta (k_+) + \bar{b} (k_+)  \Theta(-k_+)  \,,
\end{equation}
 and we get:
 \begin{equation}
  \phi (x_a)  = \int_0^{+\infty}d k_+  \frac{ e^{2 k_+}  }{ \frac 1 2 \left| e^{2 k_+} - 1 \right| } \, a (k_+) \Eos_a(k_+)  +
 \int_{-\infty}^0 d k_+  \frac{ e^{2 k_+}  }{ \frac 1 2 \left| e^{2 k_+} - 1 \right| } \, \bar{b} (k_+) \Eos_a (k_+)  
 \,,
 \end{equation} 
 where we called the on-shell waves
 \begin{equation}\label{Onshellwaves}
\Eos_a(k_+) = \E_a\left[ \omega_r(k_+) , k_+\right] \,.
\end{equation}
Notice now that $\Eos_a^\dagger (k_+) = \Eos_a(-k_+)$, so that

 \begin{equation}
\begin{aligned}
 \hat \phi (x_a)  =& \int_0^{+\infty}d k_+ \frac{ e^{2 k_+}  }{ \frac 1 2 \left| e^{2 k_+} - 1 \right| } \left[  a (k_+) \Eos_a(k_+)  +
e^{-2 k_+} \, \bar{b} (-k_+) \Eos_a^\dagger (k_+)  \right]
 \,.
\end{aligned}
 \end{equation} 
 
In the following, we will need to commute on-shell plane waves of different points. The key identity we need to calculate these commutator is Eq.~(\ref{ProductPlaneWaves}), which we reproduce here for convenience:
$$
\E_a [k] \E_b [q]   
=
e^{i \left( k_- x_a^-+ e^{-2 k_+} q_- x_b^- \right)}  e^{  i   \frac{k_+ + q_+}{1 - e^{ 2 (k_+ + q_+)}}  \left[ \left( 1 -  e^{2 k_+} \right) x^+_a +  e^{2 k_+}  \left( 1- e^{ 2 q_+} \right) x^+_b \right] } 
\,.
$$
We can then ask whether commuting two on-shell waves gives again a product  of on-shell waves, \emph{i.e.:}
\begin{equation}
\Eos_1 (k_+)
\Eos_2 (q_+)
=
\Eos_2 \left( q_+' \right)
\Eos_1 \left( k_+' \right) \,,
\end{equation}
this equation is solved by
\begin{equation}
k_+' = \frac{1}{2} \log \left(\frac{e^{2 (k_++q_+)}}{e^{2 k_+} \left(e^{2 q_+}-1\right)+1}\right)
\,,
\qquad
q_+' = \frac{1}{2} \log \left(e^{2 k_+} \left(e^{2 q_+}-1\right)+1\right) \,.
\end{equation}
We can find similar relations for Hermitian conjugate on-shell waves. The whole algebra is summarized here:
\begin{equation}\label{OnshellWavesAlgebra1}
\begin{gathered} \textstyle
\Eos_1 (k_+)\Eos_2 (q_+) =
\Eos_2 \left(  \frac{1}{2} \log \left[ e^{2 k_+} \left(e^{2 q_+}-1\right)+1 \right] \right)
\Eos_1 \left( k_++q_+ - \frac{1}{2} \log \left[ e^{2 k_+} \left(e^{2 q_+}-1\right)+1\right] \right) \,,
\\ \textstyle
\Eos_1 (k_+)\Eos_2^\dagger (q_+) =
\Eos_2^\dagger \left( q_+ - \frac 1 2 \log \left[ e^{2k_+}\left(1- e^{2q_+}\right) + e^{2q_+} \right] \right)
\Eos_1 \left( k_+ - \frac 1 2 \log \left[ e^{2k_+}\left(1- e^{2q_+} \right) + e^{2q_+} \right] \right) \,,
\\ \textstyle
\Eos_1^\dagger (k_+)\Eos_2 (q_+) =
\Eos_2 \left( \frac 1 2 \log \left[ e^{2k_+} + e^{2 q_+} -1  \right] - k_+\right)
\Eos_1^\dagger \left( \frac 1 2 \log \left[  e^{2k_+} + e^{2 q_+} -1\right] -  q_+\right) \,,
\\ \textstyle
\Eos_1^\dagger (k_+)\Eos_2^\dagger (q_+) =
\Eos_2^\dagger \left( \frac 1 2 \log \left[ 1 - e^{2q_+} \left(1- e^{2 k_+}  \right) \right] \right)
\Eos_1^\dagger \left( k_+ + q_+ -\frac 1 2 \log \left[   1 - e^{2 q_+} \left( 1- e^{2 k_+} \right) \right] \right) \,.
\end{gathered}
\end{equation}

\subsection{Two-point functions}

We are ready to study in full generality two-point functions, built from the elements of the noncommutative two-point algebra $\mathcal{A}^{\underline{\otimes}2}$ that can be written as Fourier transforms, that are $\kappa$-Poincar\'e invariant and that solve the $\kappa$-Klein--Gordon equation.
A reasonable proposal for such a function, based on what we know from commutative QFT, is something like this:
\begin{equation}
\int d^2 k  \E_1[k] \E^\dagger_2[k] \, f(k) \, \delta \left( \mathcal{C}(k) - m^2 \right) \,,
\end{equation}
where of course $f(k)$ is supposed to be a Lorentz-invariant function of the momentum. However the above function is not Lorentz-invariant. In fact:
\begin{equation}
\begin{aligned}
\int d^2 k  \E_1'[k] {\E'}^\dagger_2[k] \, f(k) \, \delta \left( \mathcal{C}(k) - m^2 \right)  
&= \int d^2 k  \E_1[\lambda(k,\omega)] \E^\dagger_2[\lambda(k,\omega)] \, f(k) \, \delta \left( \mathcal{C}(k) - m^2 \right) 
\\
&= \int d^2 q  \left| \det 
\left( \frac{\partial \lambda(k,-\omega)_\mu}{\partial k_\nu} \right) \right| \E_1[q] \E^\dagger_2[q]\, f(k) \, \delta \left( \mathcal{C}(q) - m^2 \right) 
\\
&\neq \int d^2 k  \E_1[k] \E^\dagger_2[k] \, f(k) \, \delta \left( \mathcal{C}(k) - m^2 \right) \,.
\end{aligned}
\end{equation}
Instead, inserting the square root of minus the determinant of the momentum-space metric:
\begin{equation}
\begin{aligned}
F(x^\mu_1 - x^\mu_2) &= \int d^2 k \sqrt{-g(k)}  \E_1[k] \E^\dagger_2[k] \, f(k) \, \delta \left( \mathcal{C}(k) - m^2 \right) 
\\
&= \int d^2 k \sqrt{-g(k)}  \E_1[k] \E^\dagger_2[k]  \, f(k) \, \frac{ \delta \left( k_- - \omega_r(k_+) \right) }{ \frac 1 2 \left| e^{2 k_+} - 1\right| } \,,
\end{aligned}
\end{equation}
where $ \sqrt{-g(k)} = e^{2k_+} $, makes the integral Lorentz invariant.

Now we worry about another issue: ordering dependence. We could have used the Weyl-ordered basis of plane waves to construct the function:
\begin{equation}
\int d^2 q \sqrt{-g'(q)}  \F_1 [q] \F^\dagger_2 [q] \, f'(q) \, \delta \left( \mathcal{C'}(q) - m^2 \right) 
= \int d^2 q \sqrt{-g'(q)}  \F_1 [q] \F^\dagger_2 [q] \, f'(q)\, \frac{ \delta \left( q_- - \omega_w(q_+) \right) }{\left| \frac{\left( e^{2 q_+} - 1 \right)^2}{4 q_+}\right| } \,,
\end{equation}
where $g'= \left| \frac{e^{2q_+} -1}{2q_+} \right|$, $\mathcal{C'}(q)= \mathcal{C} \left[ q_+,q_- \left( \frac{e^{2 q_+} -1}{2 q_+} \right) \right]$ , $f'(q)=f\left[ q_+,q_- \left( \frac{e^{2 q_+} -1}{2 q_+} \right) \right]$. However, we can prove that the two functions are identical. In fact, under the coordinate change $q_+ = k_+$, $q_- = \frac{2 k_+ k_-}{e^{2 k_+} -1}$, one has:
\begin{equation}
d^2 q \sqrt{-g'(q)} =  d^2 k  \sqrt{-g(k)} \,, ~~
\F_1[q] = \E_1[k] \,, ~~
\F^\dagger_2[q] = \E^\dagger_2[k] \,, \qquad h'(q)=h(k) \,, ~~
\mathcal{C}'(q) = \mathcal{C}(k) \,,
\end{equation}
and therefore
\begin{equation}
\begin{aligned}
\int d^2 q \sqrt{-g'}  \F_1 [q] \F^\dagger_2 [q] \, f'(q) \, \delta \left( \mathcal{C'}(q) - m^2 \right) 
&= \int d^2 k \sqrt{-g}  \E_1 [k] \E^\dagger_2 [k] f(k) \frac{ \delta \left( \frac{2 k_+ k_-}{e^{2 k_+} -1} - \omega_w(k_+) \right) }{\left| \frac{\left( e^{2 k_+} - 1 \right)^2}{4k_+}\right| }
\\
&= \int d^2 k \sqrt{-g)}  \E_1 [k] \E^\dagger_2 [k] f(k) \frac{ \delta \left(k_- -  \frac{e^{2 k_+} -1}{2 k_+ } \omega_w(k_+) \right) }{\left|\frac{2 k_+ }{e^{2 k_+} -1} \right|\left| \frac{\left( e^{2 k_+} - 1 \right)^2}{4k_+}\right| }
\\
&= \int d^2 k \sqrt{-g}  \E_1[k] \E^\dagger_2[k] f(k) \frac{ \delta \left( k_- - \omega_r(k_+) \right) }{ \frac 1 2 \left| e^{2 k_+} - 1\right| }   \,.
\end{aligned}
\end{equation}

The function $f(k)$ appearing in our two-point function should be Lorentz-invariant, and the functions that are used for commutative QFT two-point functions, \emph{e.g.} Feynmann propagators, Wightman  functions and Pauli-Jordan functions, are all constants on the forward and backward light cones in momentum space. In our case we can write:
\begin{equation}
f(k) = f_- \, \Theta(-k_+) + f_+ \, \Theta(k_+) \,,
\label{f1f2}
\end{equation} 
where $f_-$ and $f_+$ are constants. This function gives $f_+$ on the forward light cone and $f_-$ on the backwards one, and it is easy to see that it is Lorentz invariant, because the sign of $k_+$ is not changed by on-shell Lorentz transformations. This expression, however is not \emph{globally} Lorentz-covariant: the backwards light cone is not closed under Lorentz transformations, and this will make the $f_-$ term non-invariant. Let us now calculate explicitly the form of $F(x^\mu_1 - x^\mu_2)$ that is implied by the choice~(\ref{f1f2}):
\begin{equation}
\begin{aligned}
F(x^\mu_1 - x^\mu_2) &= \int_\mathbbm{R} d k_+ \frac{e^{2 k_+}}{ \frac 1 2 \left| e^{2 k_+} - 1\right| } e^{i \left( \frac{2 m^2}{e^{2k_+} -1 } \right) \left(  x_1^- - x_2^- \right)} e^{i \left(\frac{e^{2 k_+}-1}{2}\right) \left(x_1^+-x_2^+\right) }  f \left(k_+,\frac{2 m^2}{e^{2k_+} -1 } \right)   
\\
&= \int_0^\infty  \frac{d y}{\left| y - 1\right| } e^{i \left( \frac{2 m^2}{y -1 } \right) \left(  x_1^- - x_2^- \right)} e^{i \left(\frac{y-1}{2}\right) \left(x_1^+-x_2^+\right) }  f \left(\frac 1 2 \log y,\frac{2 m^2}{y -1 } \right)   
\\
&= \int_{-1}^\infty  \frac{d z}{\left| z \right| } e^{i \left( \frac{2 m^2}{z} \right) \left(  x_1^- - x_2^- \right)} e^{i \left(\frac{z}{2}\right) \left(x_1^+-x_2^+\right) }  f \left(\frac 1 2 \log (z+1),\frac{2 m^2}{z} \right)   
\\
&= f_- \int_0^1 \frac{d u}{u } e^{- i \left( \frac{2 m^2}{u} \right) \left(  x_1^- - x_2^- \right)} e^{-i \left(\frac{u}{2}\right) \left(x_1^+-x_2^+\right) }   
+ f_+ \int_0^\infty  \frac{d z}{ z  } e^{i \left( \frac{2 m^2}{z} \right) \left(  x_1^- - x_2^- \right)} e^{i \left(\frac{z}{2}\right) \left(x_1^+-x_2^+\right) }   
\,,
\end{aligned}
\end{equation}
Reintroducing $\kappa$, the expression above becomes:
\begin{equation}
\begin{aligned}
F(x^\mu_1 - x^\mu_2) =& f_- \int_0^1 \frac{d u}{u } e^{- i \left( \frac{2 m^2}{\kappa \, u} \right) \left(  x_1^- - x_2^- \right)} e^{-i \left(\frac{\kappa  \,u}{2}\right) \left(x_1^+-x_2^+\right) }   
+ f_+ \int_0^\infty  \frac{d z}{ z  } e^{i \left( \frac{2 m^2}{\kappa \, z} \right) \left(  x_1^- - x_2^- \right)} e^{i \left(\frac{\kappa \, z}{2}\right) \left(x_1^+-x_2^+\right) }   
\\
=& f_- \int_0^{\frac \kappa {2 m}} \frac{d u}{u } e^{- i m \left( \frac{1}{u} \right) \left(  x_1^- - x_2^- \right)} e^{-i m u \left(x_1^+-x_2^+\right) }   
+ f_+ \int_0^\infty  \frac{d z}{ z  } e^{ i m \left( \frac{1}{u} \right) \left(  x_1^- - x_2^- \right)} e^{i m u \left(x_1^+-x_2^+\right) }  
\\
=& f_- \int_{-\infty}^{\log \frac \kappa {2 m}} d \chi e^{- i m \left( \cosh \chi - \sinh \chi \right) \left(  x_1^- - x_2^- \right)} e^{-i m \left( \cosh \chi + \sinh \chi \right) \left(x_1^+-x_2^+\right) } 
\\&~  
+ f_+ \int_{-\infty}^\infty  d \chi e^{ i m  \left( \cosh \chi - \sinh \chi \right)\left(  x_1^- - x_2^- \right)} e^{i m  \left( \cosh \chi + \sinh \chi \right) \left(x_1^+-x_2^+\right) }  
\\
=& f_- \int_{-\infty}^{m \sinh \left(\log \frac \kappa {2 m}\right)} \frac{dp}{\sqrt{p^2 + m^2}} e^{- i   \left( \sqrt{p^2 + m^2} - p \right)\left(  x_1^- - x_2^- \right) -i   \left(  \sqrt{p^2 + m^2} + p  \right) \left(x_1^+-x_2^+\right) }  
\\&~  
+ f_+ \int_{-\infty}^\infty  \frac{dp}{\sqrt{p^2 + m^2}} e^{i   \left( \sqrt{p^2 + m^2} - p \right)\left(  x_1^- - x_2^- \right) + i  \left(  \sqrt{p^2 + m^2} + p  \right) \left(x_1^+-x_2^+\right) }  
\\
=& f_- \int_{-\infty}^{m \sinh \left(\log \frac \kappa {2 m}\right)} \frac{dp}{\sqrt{p^2 + m^2}} e^{- 2 i   \left[ \sqrt{p^2 + m^2} ( x^0_1 - x^0_2) + p ( x^1_1 - x^1_2) \right]}
\\&~  
+ f_+ \int_{-\infty}^\infty  \frac{dp}{\sqrt{p^2 + m^2}}  e^{2 i   \left[ \sqrt{p^2 + m^2} ( x^0_1 - x^0_2) + p ( x^1_1 - x^1_2) \right]}
\,,
\end{aligned}
\end{equation}
the expression above is identical to the integrals appearing in the undeformed 2-point functions (written in light-cone coordinates), except for the Lorentz-breaking integration boundary $m \sinh \left(\log \frac \kappa {2 m}\right)$ in the first integral.

So, our conclusion is that, in order to have a $\kappa$-Poincar\'e-invariant function of type $F(x^\mu_1 - x^\mu_2) $, we have to set $f_- =0$. We have found a first $\kappa$-Poincar\'e-invariant two-point function, based on the translation-invariant wave combination~(\ref{TranslationInvariantPlaneWave1}), $\E_1 [k] \E_2^\dagger [k]$:
\begin{equation}
F = \int d^2 k \sqrt{-g(k)}  \E_1[k] \E^\dagger_2[k] \, \Theta(k_+) \, \delta \left( \mathcal{C}(k) - m^2 \right) =  \int_{-\infty}^\infty  \frac{dp}{\sqrt{p^2 + m^2}}  e^{2 i   \left[ \sqrt{p^2 + m^2} ( x^0_1 - x^0_2) + p ( x^1_1 - x^1_2) \right]} \,.
\end{equation}
We could have instead used the translation invariant combination of plane waves introduced in Eq.~(\ref{TranslationInvariantPlaneWave3}), $\E_2 [k] \E_1^\dagger [k]$, but this is just the Hermitian conjugate of the wave combination used before. Moreover, the two-point function built with it coincides with $F$ with $x^\mu_1$ and $x^\mu_2$ exchanged, because:
\begin{equation}
F^\dagger(x^\mu_1 - x^\mu_2) = F(x^\mu_2 - x^\mu_1) \,.
\end{equation}

The wave combination~(\ref{TranslationInvariantPlaneWave2}), $\E_1^\dagger [k] \E_2 [k]$ is not obviously related to~(\ref{TranslationInvariantPlaneWave1}), so we need to check what we get if we use it to define our two-point function:
\begin{equation}
H(x^\mu_1 - x^\mu_2) = \int d^2 k  \E^\dagger_1[k] \E_2[k] \, h(k) \, \delta \left( \mathcal{C}(k) - m^2 \right)  \,,
\end{equation}
which is Lorentz-invariant because, from Eq.~(\ref{TransformationLawChiWaves}),  $ \E^\dagger_1[k] \E_2[k]  =  \E^\dagger_1[\lambda(k,\omega \triangleleft S[k])] \E_2[\lambda(k,\omega \triangleleft S[k])] $, and the Jacobian of the transformation $q_\mu = \lambda_\mu(k,\omega \triangleleft S[k])$ is one.

Again, the plane wave combination~(\ref{TranslationInvariantPlaneWave4}), $\E_2^\dagger [k] \E_1 [k]$ is just the Hermitian conjugate of~(\ref{TranslationInvariantPlaneWave2}), and again, the two-point function built with it coincides with $H$ with $x^\mu_1$ and $x^\mu_2$ exchanged, because:
\begin{equation}
H^\dagger(x^\mu_1 - x^\mu_2) = H(x^\mu_2 - x^\mu_1) \,.
\end{equation}

An explicit calculation of $H$ gives:
\begin{equation}
\begin{aligned}
H(x^\mu_1 - x^\mu_2) &= \int_\mathbbm{R} d k_+ \frac{1}{ \frac 1 2 \left| e^{2 k_+} - 1\right| } e^{- i e^{2 k_+}\left( \frac{2 m^2}{e^{2k_+} -1 } \right) \left(  x_1^- - x_2^- \right)} e^{- i \left(\frac{1- e^{-2 k_+}}{2}\right) \left(x_1^+-x_2^+\right) }  h \left(k_+,\frac{2 m^2}{e^{2k_+} -1 } \right)   
\\
&= \int_\mathbbm{R} d k_+ \frac{ e^{-2 k_+} }{ \frac 1 2 \left| e^{-2 k_+} - 1\right| } e^{ i \left( \frac{2 m^2}{e^{-2k_+} -1 } \right) \left(  x_1^- - x_2^- \right)} e^{ i \left(\frac{e^{-2 k_+}-1}{2}\right) \left(x_1^+-x_2^+\right) } h \left(k_+,\frac{2 m^2}{e^{2k_+} -1 } \right)   
\\
&= \int_0^\infty  \frac{d y}{\left| y - 1\right| } e^{i \left( \frac{2 m^2}{y -1 } \right) \left(  x_1^- - x_2^- \right)} e^{i \left(\frac{y-1}{2}\right) \left(x_1^+-x_2^+\right) } h \left(-\frac 1 2 \log y,\frac{2 m^2}{y -1 } \right)   
\\
&= \int_{-1}^\infty  \frac{d z}{\left| z \right| } e^{i \left( \frac{2 m^2}{z} \right) \left(  x_1^- - x_2^- \right)} e^{i \left(\frac{z}{2}\right) \left(x_1^+-x_2^+\right) } h \left(-\frac 1 2 \log (z+1),\frac{2 m^2}{z} \right)   
\,, 
\end{aligned}
\end{equation}
and, if $h(k) = h_- \Theta(- k_+) + h_+ \Theta (k_+)$:
\begin{equation}
\begin{aligned}
H(x^\mu_1 - x^\mu_2) &= h_+ \int_0^1 \frac{d u}{u } e^{- i \left( \frac{2 m^2}{u} \right) \left(  x_1^- - x_2^- \right)} e^{-i \left(\frac{u}{2}\right) \left(x_1^+-x_2^+\right) }   
+ h_- \int_0^\infty  \frac{d z}{ z  } e^{i \left( \frac{2 m^2}{z} \right) \left(  x_1^- - x_2^- \right)} e^{i \left(\frac{z}{2}\right) \left(x_1^+-x_2^+\right) }  
\\
=& h_+ \int_{-\infty}^{m \sinh \left(\log \frac \kappa {2 m}\right)} \frac{dp}{\sqrt{p^2 + m^2}} e^{- 2 i   \left[ \sqrt{p^2 + m^2} ( x^0_1 - x^0_2) + p ( x^1_1 - x^1_2) \right]}
\\&~  
+ h_- \int_{-\infty}^\infty  \frac{dp}{\sqrt{p^2 + m^2}}  e^{2 i   \left[ \sqrt{p^2 + m^2} ( x^0_1 - x^0_2) + p ( x^1_1 - x^1_2) \right]}
\,, 
\,, 
\end{aligned}
\end{equation}
so, if we set $h_+=0$ we have a genuinely Lorentz-invariant function. This function, however, turns out to be identical to $F$ (modulo a constant factor).

We conclude that we can use $F(x^\mu_1 - x^\mu_2)$ and its Hermitian conjugate to define all two-point functions that we need, which will have the appropriate commutative limit and invariance properties. Moreover, these two-point functions will be indistinguishable from their commutative counterparts. For example, the Wightman function can be defined as:
\begin{equation}
\Delta_\text{W} (x^\mu_1 - x^\mu_2) = 
\int d^2 k \sqrt{-g(k)}  \, \E_1[k] \E^\dagger_2[k] \,  \Theta(k_+) \, \delta \left( \mathcal{C}(k) - m^2 \right)  \,,
\end{equation}
and the associated Pauli-Jordan function will be the anti-Hermitian part of $\Delta_\text{W}$:
\begin{equation}
\Delta_\text{PJ} (x^\mu_1 - x^\mu_2) =  
\int d^2 k \sqrt{-g(k)}  \left( \E_1[k] \E^\dagger_2[k]  - \E_2[k] \E^\dagger_1[k] \right) \Theta(k_+) \, \delta \left( \mathcal{C}(k) - m^2 \right)  \,.
\end{equation}

\subsection{Field quantization}

We can use the Pauli-Jordan  function to define a quantization, \emph{i.e.}
\begin{equation}
[ \hat \phi (x_1) , \hat \phi^\dagger (x_2) ] = i \Delta_\text{PJ} (x^\mu_1 - x^\mu_2) \,,
~~
[ \hat \phi (x_1) , \hat \phi (x_2) ] = 0 \,,
~~
[ \hat \phi^\dagger (x_1) , \hat \phi^\dagger (x_2) ] = 0\,,
\end{equation}
where now the Fourier coefficients of our on-shell field are assumed to be non-necessarily commutative operators, which however commute with $x^\mu_a$:
 \begin{equation}
  \hat \phi (x_a)  = \int_0^{+\infty}d k_+  \frac{ e^{2 k_+}  }{ \frac 1 2 \left| e^{2 k_+} - 1 \right| } \left( \hat a (k_+) \Eos_a(k_+)  +
  e^{-2 k_+}    \, \hat{b}^\dagger (k_+) \Eos_a^\dagger (k_+)  \right)
 \,,
 \end{equation} 
 and the Hermitian conjugate field will be:
  \begin{equation}
  \hat \phi^\dagger (x_a)  = \int_0^{+\infty}d k_+  \frac{ e^{2 k_+}  }{ \frac 1 2 \left| e^{2 k_+} - 1 \right| } \left( \hat a^\dagger (k_+) \Eos_a^\dagger(k_+)  +
  e^{-2 k_+}    \, \hat{b} (k_+) \Eos_a^ (k_+)  \right)
 \,.
 \end{equation} 

Consider first the equation $[ \hat \phi (x_1) , \hat \phi^\dagger (x_2) ] = i \Delta_\text{PJ} (x^\mu_1 - x^\mu_2)$, which implies:
\begin{equation}\label{QuantizationRules1}
\begin{aligned}
&\int_0^\infty \int_0^\infty dk_+ dq_+  \frac{  e^{2 (k_+ + q_+)}  }{ \frac 1 4 \left| e^{2 k_+} - 1  \right|\left| e^{2 q_+} - 1  \right| } \left\{    \hat a (k_+) \hat a^\dagger (q_+) \Eos_1(k_+)  \Eos_2^\dagger (q_+)  
 -  \hat a^\dagger (q_+)\hat a (k_+) ~ \Eos_2^\dagger (q_+)   \Eos_1(k_+)   \right\} 
 \\
& = \int_0^\infty  dk_+   \frac{ e^{2 k_+}  }{ \frac 1 2 \left| e^{2 k_+} - 1 \right| }  \Eos_1(k_+)  \Eos_2^\dagger (k_+)  \,,
\\
&\int_0^\infty  \int_0^\infty dk_+ dq_+  \frac{  e^{2  k_+ }  }{ \frac 1 4 \left| e^{2 k_+} - 1  \right|\left| e^{2 q_+} - 1  \right| } \left\{      \hat a (k_+)  \hat b  (q_+)  \, \Eos_1(k_+) \Eos_2(q_+)    - \hat b  (q_+) \hat a (k_+) \, \Eos_2(q_+) \Eos_1(k_+)   \right\}  = 0 \,,
\\
&\int_0^\infty \int_0^\infty  dk_+ dq_+  \frac{  e^{2  q_+ }  }{ \frac 1 4 \left| e^{2 k_+} - 1  \right|\left| e^{2 q_+} - 1  \right| }  \left\{   \hat b^\dagger (k_+) \hat a^\dagger (q_+) \,  \Eos_1^\dagger (k_+)  \Eos_2^\dagger (q_+)  - 
 \hat a^\dagger (q_+)\hat b^\dagger (k_+) \, \Eos_2^\dagger (q_+) \Eos_1^\dagger (k_+)    \right\} = 0 \,,
\\
&\int_0^\infty \int_0^\infty  dk_+ dq_+  \frac{  1 }{ \frac 1 4 \left| e^{2 k_+} - 1  \right|\left| e^{2 q_+} - 1  \right| }   \left\{     \hat b^\dagger (k_+) \hat b  (q_+) \,\Eos_1^\dagger (k_+)   \Eos_2(q_+)   
-   \hat b  (q_+)\hat b^\dagger (k_+)  \, \Eos_2(q_+)   \Eos_1^\dagger (k_+)  \right\}  
\\&=  - \int_0^\infty  dk_+  \frac{ e^{2 k_+}  }{ \frac 1 2 \left| e^{2 k_+} - 1 \right| } \Eos_2 (k_+)  \Eos_1^\dagger(k_+) \,,
\end{aligned}
\end{equation}
recall Eq.~(\ref{OnshellWavesAlgebra1}), and rewrite it according to our present needs:
\begin{equation}\label{OnshellWavesAlgebra2}
\begin{gathered} \textstyle
\Eos_2^\dagger (q_+)\Eos_1 (k_+) =
\Eos_1 \left( \frac 1 2 \log \left[ e^{2q_+} + e^{2 k_+} -1  \right] - q_+\right)
\Eos_2^\dagger \left( \frac 1 2 \log \left[  e^{2q_+} + e^{2 k_+} -1\right] - k_+\right) \,,
\\\textstyle
\Eos_2 (q_+)\Eos_1 (k_+) =
\Eos_1 \left(  \frac{1}{2} \log \left[ e^{2 q_+} \left(e^{2 k_+}-1\right)+1 \right] \right)
\Eos_2 \left( q_++k_+ - \frac{1}{2} \log \left[ e^{2 q_+} \left(e^{2 k_+}-1\right)+1\right] \right) \,,
\\\textstyle
\Eos_1^\dagger (k_+)\Eos_2^\dagger (q_+) =
\Eos_2^\dagger \left( \frac 1 2 \log \left[ 1 - e^{2q_+} \left(1- e^{2 k_+}  \right) \right] \right)
\Eos_1^\dagger \left( k_+ + q_+ -\frac 1 2 \log \left[   1 - e^{2 q_+} \left( 1- e^{2 k_+} \right) \right] \right)  \,,
\\\textstyle
\Eos_1^\dagger (k_+)\Eos_2 (q_+) =
\Eos_2 \left( \frac 1 2 \log \left[ e^{2k_+} + e^{2 q_+} -1  \right] -k_+\right)
\Eos_1^\dagger \left( \frac 1 2 \log \left[  e^{2k_+} + e^{2 q_+} -1\right] - q_+\right) \,.
\end{gathered}
\end{equation}
Consider the first line of Eq.~(\ref{QuantizationRules1}). Using~(\ref{OnshellWavesAlgebra2}), we can rewrite it as
\begin{equation}\label{PassageOscillatorAlgebra1}
\begin{aligned}
&\int_{\mathbbm{R}_+^2} dk_+ dq_+  \frac{  e^{2 (k_+ + q_+)}  }{ \frac 1 4 \left| e^{2 k_+} - 1  \right|\left| e^{2 q_+} - 1  \right| } \Bigg{\{}    \hat a (k_+) \hat a^\dagger (q_+) \Eos_1(k_+)  \Eos_2^\dagger (q_+)  
\\&
 -  \hat a^\dagger (q_+)\hat a (k_+) ~ \Eos_1 \left( \frac 1 2 \log \left[ e^{2q_+} + e^{2 k_+} -1  \right] - q_+\right)
\Eos_2^\dagger \left( \frac 1 2 \log \left[  e^{2q_+} + e^{2 k_+} -1\right] - k_+\right)  \Bigg{\}} \,,
\end{aligned}
\end{equation}
and then, inverting the relations
\begin{equation}\label{PassageOscillatorAlgebra2}
    k_+' =  \frac 1 2 \log \left[ e^{2q_+} + e^{2 k_+} -1  \right] - q_+
\,,
\qquad q_+' = \frac 1 2 \log \left[  e^{2q_+} + e^{2 k_+} -1\right] - k_+   \,,
\end{equation}
we get:
\begin{equation}\label{PassageOscillatorAlgebra3}
q_+ = q'_+ - \frac 1 2 \log \left[ e^{2 k'_+} + e^{2 q'_+} - e^{2(k'_+ + q'_+)} \right] \,,
\qquad
k_+ =  k'_+ - \frac 1 2 \log \left[ e^{2 k'_+} + e^{2 q'_+} - e^{2(k'_+ + q'_+)} \right] \,,
\end{equation}
and, taking into account the Jacobian of the transformation $\left|  e^{-2k_+'} + e^{-2 1_+'} -1\right|^{-1}$,
\begin{equation}
    \begin{aligned}
    &\int_{\mathbbm{R}_+^2} \frac{ dk_+ dq_+    e^{2 (k_+ + q_+)}  }{ \frac 1 4 \left| e^{2 k_+} - 1  \right|\left| e^{2 q_+} - 1  \right| }  \Eos_1(k_+)  \Eos_2^\dagger (q_+) \Bigg{\{}    \hat a (k_+) \hat a^\dagger (q_+)    - \frac{ \frac 1 2 \left| e^{2 k_+} - 1 \right| }{ e^{2 k_+}  } \delta( k_+ - q_+) 
    \\&
 - \frac{\hat a^\dagger \left( q_+ - \frac 1 2 \log \left[ e^{2 k_+} + e^{2 q_+} - e^{2(k_+ + q_+)} \right] \right) \hat a \left( k_+ - \frac 1 2 \log \left[ e^{2 k_+} + e^{2 q_+} - e^{2(k_+ + q_+)} \right] \right) }{\left| 1- e^{-2 k_+} - e^{-2q_+} \right| }  \Bigg{\}}  
 = 0 \,,
    \end{aligned}
\end{equation}
which imposes the following deformed commutators for the creation and annihilation operators:
\begin{equation}
    \begin{aligned}
   \textstyle  & \hat a (k_+) \hat a^\dagger (q_+)    - \frac{\hat a^\dagger \left( q_+ - \frac 1 2 \log \left[ e^{2 k_+} + e^{2 q_+} - e^{2(k_+ + q_+)} \right] \right) \hat a \left( k_+ - \frac 1 2 \log \left[ e^{2 k_+} + e^{2 q_+} - e^{2(k_+ + q_+)} \right]  \right) }{\left| 1- e^{-2 k_+} - e^{-2q_+} \right|} =
   \\&  \frac 1 2 \left| 1- e^{-2 k_+} - 1 \right| \delta( k_+ - q_+) \,.
    \end{aligned}
\end{equation}
All the other commutators forming the bosonic oscillator algebra can be similarly derived.

Notice now that, upon commuting $ \hat a (k_+) $ and $\hat a^\dagger (q_+)   $, we get creation and annihilation operators labeled by momentum coordinates that diverge, or become complex, for certain values of $k_+$ and $q_+$:
\begin{equation}
    q_+'' = q_+ - \frac 1 2 \log \left[ e^{2 k_+} + e^{2 q_+} - e^{2(k_+ + q_+)} \right] \,, \qquad  
    k_+'' = k_+ - \frac 1 2 \log \left[ e^{2 k_+} + e^{2 q_+} - e^{2(k_+ + q_+)} \right] \,,
\end{equation}
when $e^{2 k_+} = \frac{1}{1- e^{-2 q_+}}$ both $q_+''$ and $k_+''$ diverge.
This has to do with the fact that the maps that send the momenta of the on-shell waves in Eq.~(\ref{OnshellWavesAlgebra1}) to the momenta of the commuted waves are not maps of $\mathbbm{R}_+^2$ onto itself.  Specifically, when we reached Eq.~(\ref{PassageOscillatorAlgebra1}), we had to make the coordinate transformation~(\ref{PassageOscillatorAlgebra3}), which, as a real map, sends the region
\begin{equation}
    e^{2 k'_+} > \frac{1}{1- e^{-2q_+'}} \,,
\end{equation}
into the region
\begin{equation}
    e^{2k_+} > 1 -e^{2 q_+} \,.
\end{equation}
We need to consider what happens beyond those regions, which can be accessed by Lorentz-transforming the momenta, and cannot therefore be ignored if we want to preserve Lorentz invariance.  This issue deserves further investigation.

\section{Conclusions}

We solved the main problem that obstructed the definition of a genuine $\kappa$-Poincar\'e-invariant QFT on $\kappa$-Minkowski, defined in terms of ``noncommutative'' N-point functions. This was the problem of defining in a  $\kappa$-Poincar\'e-covariant way the algebra of functions of more than one point, which we called $\mathcal{A}^{\bar{\otimes}N}$. We did this at the expense of generality: a covariant algebra can be defined only for the ``lightlike'' $\kappa$-Minkowski algebra $v^\mu v^\nu g_{\mu\nu} = 0$. 

We introduced a natural representation of the algebra $\mathcal{A}^{\bar{\otimes}N}$, and found that translation-invariant coordinate differences belong to the maximal Abelian subalgebra of $\mathcal{A}^{\bar{\otimes}N}$, and therefore they are, for all practical purposes, equivalent to commutative functions.

This result has a consequence that hugely simplifies the interpretational framework of the QFT: all N-point functions are translation-invariant, and they are therefore commutative. A QFT on $\kappa$-Minkowski  can then be defined in terms of a set of standard $N$-point functions, just like any QFT on the ordinary, commutative Minkowski space.

We studied explicitly the possible 2-point functions, defined by requiring that they solve the $\kappa$-Klein--Gordon equation and that they are $\kappa$-Poincar\'e invariant. This gives a Wightman function that is equivalent to the commutative one, with all the dependence on the deformation parameter $\kappa$ disappearing from the theory. All $2$-point functions that can be built from it, like the Pauli--Jordan function, will be therefore undeformed and independent of $\kappa$.

With the Pauli--Jordan function, we can impose quantization rules for free complex $\kappa$-Klein--Gordon fields, and look for a representation of the quantum fields in terms of a bosonic oscillator algebra. One finds that the algebra of bosonic oscillators is deformed, similarly to other results in the $\kappa$-QFT literature~(\emph{e.g.} \cite{Daszkiewicz:2007az,Arzano2007}). However, the commutation relations of our creation and annihilation operators seem to involve divergent/complex momenta, an issue whose investigation we leave to future works.

The fact that our $2$-point functions are undeformed motivates the conjecture that all $N$-point functions of the free theory might turn out to be undeformed and independent of $\kappa$, which would make the theory completely indistinguishable from the ordinary, commutative free scalar QFT on Minkowski space. Indeed, this is what happened in~\cite{Oeckl:2000eg,Wess:2003da,Chaichian:2004za,Koch:2004ud,Aschieri:2005zs,Fiore:2007vg,Fiore:2007zz} (see in particular~\cite{Fiore:2007zz}) for the free scalar QFT on the Moyal--Weyl noncommutative spacetime. In these works, extending the noncommutative algebra of coordinates to a deformed tensor product algebra which is covariant under noncommutative Poincar\'e transformations, resulted in a mostly-commutative algebra, in which all translation-invariant coordinate differences are commutative, just like our result. Both the free and the interacting scalar QFT turns out to be equivalent to the commutative/undeformed one~\cite{Fiore:2007vg,Fiore:2007zz}. We proved a similar result only for the free theory, and only for $2$-point functions. One of the first priorities for further works in this direction will be to investigate whether the same holds for all $N$-point functions in the free theory, which seems likely. Then, the following step will be to investigate an interacting theory, and check whether a dependence on $\kappa$ finally appears in interaction vertices.

Another interesting issue is the relation of our construction with the approaches based on star products~\cite{Kosinski1999,Kosinski2001,Kosinski:2003xx,Arzano2007,Freidel:2007hk,Arzano:2009ci,Arzano:2017uuh,Juric2018,Arzano2018,Poulain2018,Mathieu2020}. In particular,~\cite{Juric2015} focuses on the lightlike $\kappa$-Minkowski spacetime, and, despite being based on a star-product approach whose fundamental ontology is that of commutative functions, it derives some results that are in line with ours so far: the free scalar QFT is undeformed, and a dependence on $\kappa$ seems to be confined to the interacting theory. An approach based on star products isn't obviously related to ours, based on a covariant braided $N$-point algebra of coordinates, but it would be very interesting if one could prove a relation between the two. In the case of QFT on the Moyal noncommutative spacetime, the two approaches are fundamentally different and lead to different predictions for the $N$-point functions~\cite{Fiore:2007vg}.

We have shown how to have, for a free two-point function a $\kappa$-Poincar\'e invariant \emph{on shell} theory, by entirely avoiding the Lorentz-breaking parts of the mass shell. This workaround might not work in the interacting theory, which requires loop integrations of off-shell momenta. If these parts of momentum space cannot be avoided, perhaps a breaking of Lorentz symmetry can be avoided by incorporating into our theory the plane waves that are obtained by boosting the waves belonging to the ``Lorentz-breaking'' mass-shell beyond the patch of momentum space that is covered by our coordinates. Then, as can be seen in  relation~\eqref{LorentzTransformMomentum} and the like, one gets logarithms of negative numbers, \emph{i.e.} complex frequencies. This might indicate some sort of damping, and deserves further scrutiny.

\appendix

\section{Appendix: some $\kappa$-Minkowski algebraic calculations}
\label{AppendixKappaAlgebra}

In this appendix we explicitly derive some useful identities. We explicitly reintroduce $\kappa$, as it will be expedient in some cases to have it be a different constant.
We start with the commutation rules
\begin{equation}
[T , X] = \frac{i}{\kappa} X \,,
\end{equation}
can be used repeatedly to prove inductively that
\begin{equation}
\begin{gathered}
T \, X = X (T + i/\kappa) \,,
\\
T^2 \, X = X (T + i/\kappa)^2 \,,
\\
\vdots
\\
T^n \, X = X (T + i/\kappa)^n \,,
\end{gathered}
\end{equation}
therefore
\begin{equation}
e^{i p_0 T} \, X= \sum_{n=0}^\infty \frac{(i p_0)^n}{n!} T^n \, X
= X \sum_{n=0}^\infty \frac{(i p_0)^n}{n!} (T + i/\kappa)^n = X \, e^{i p_0 T - p_0/\kappa } \,,
\end{equation}
and
\begin{equation}
\begin{gathered}
e^{i p_0 T} \, X = e^{-p_0/\kappa} X \, e^{i p_0 T} \, \,,
\\
e^{i p_0 T} \, X^2 = e^{-2 p_0/\kappa} X^2 \, e^{i p_0 T}  \,,
\\
\vdots
\\
e^{i p_0 T} \, X^n = e^{-n p_0/\kappa} X^n \, e^{i p_0 T}  \,,
\end{gathered}
\end{equation}
so we conclude that
\begin{equation}\label{CommutatorExpTExpX}
e^{i p_0 T}  e^{i p_1 X} = e^{i p_0 T} \sum_{n=0}^\infty \frac{(i p_1)^n}{n!} X^n 
=
  \sum_{n=0}^\infty \frac{(i p_1)^n}{n!} e^{-n p_0/\kappa} X^n \, e^{i p_0 T} 
= e^{i e^{-p_0/\kappa} p_1 X} e^{i p_0 T} \,.
\end{equation}

The product of two right-ordered plane waves is then
\begin{equation}\label{RightOrderedCombinationLaw}
 e^{i p_1 X}  e^{i p_0 T} e^{i q_1 X}   e^{i q_0 T}  
=e^{i p_1 X}   e^{i e^{-p_0/\kappa} q_1 X} e^{i p_0 T}  e^{i q_0 T}  =e^{i (p_1 + e^{-p_0/\kappa} q_1 ) X}  e^{i (p_0+q_0) T}  
 \,.
\end{equation}
similarly, left-ordered plane waves combine in the following way:
\begin{equation}
 e^{i p_0 T}  e^{i p_1 X}  e^{i q_0 T}  e^{i q_1 X}  
= e^{i (p_0+q_0) T}  e^{i (e^{+q_0/\kappa} p_1 +  q_1 ) X} 
 \,.
\end{equation}

Weyl-ordered waves are a little bit more tricky. First we need to find their relation with right-ordered waves. To do so, expand Eq.~(\ref{CommutatorExpTExpX}) to first order in $p_0$:
\begin{equation}
e^{i p_1 X} T =    \left( T + \frac{p_1}{\kappa} X \right)  e^{i p_1 X}  \,,
\end{equation}
by induction,
\begin{equation}
e^{i p_1 X} T^n =    \left( T + \frac{p_1}{\kappa} X \right)^n  e^{i p_1 X}  \,,
\end{equation}
and so
\begin{equation}
e^{i p_1 X} e^{ i p_0 T} =   e^{ i p_0 \left( T + \frac{p_1}{\kappa} X \right)}  e^{i p_1 X}  \,.
\end{equation}
Multiply now both sides by $e^{-i p_1 X}$ from the right, and reorder the left hand side with $T$ to the right:
\begin{equation}
\begin{gathered}
e^{i p_1 X} e^{ i p_0 T} e^{-i p_1 X}=   e^{ i p_0 \left( T + \frac{p_1}{\kappa} X \right)}   \,,
\\
e^{i\left(1 - e^{- p_0/\kappa}\right) p_1 X}e^{ i p_0 T} =   e^{ i p_0 \left( T + \frac{p_1} {\kappa} X \right)}   \,,
\end{gathered}
\end{equation}
if we now rename $p_0 = q_0$ and $\frac{p_1 p_0} {\kappa} = q_1$ we get the desired expression:
\begin{equation}\label{WeylOrderedCombinationLaw0}
e^{ i \left( q_0  T + q_1 X \right)}   =
e^{i\left(\frac{1 - e^{- q_0/\kappa}}{q_0/\kappa}\right)  q_1 X } e^{ i q_0 T}
\,.
\end{equation}

Now that we know how to translate Weyl-ordered waves into right-ordered ones, we can use the combination law of the latters to derive the one of the formers. Consider, in fact, the following rewriting of Eq.~(\ref{RightOrderedCombinationLaw}):
\begin{equation}
 e^{i \left(\frac{1 - e^{- \frac{p_0}{\kappa}}}{p_0 /\kappa}\right) p_1 X}  e^{i p_0 T} e^{i \left(\frac{1 - e^{- \frac{q_0}{\kappa}}}{q_0 / \kappa}\right)  q_1 X}   e^{i q_0 T}  
=
e^{i \left[ \left(\frac{1 - e^{- \frac{p_0}{\kappa}}}{p_0/\kappa}\right)  p_1 + e^{-\frac{p_0}{\kappa}} \left(\frac{1 - e^{- \frac{q_0}{\kappa}}}{q_0 / \kappa}\right)  q_1 \right] X}  e^{i (p_0+q_0) T}  
 \,,
\end{equation}
converting the right-hand side into a Weyl-ordered wave through the inverse relation to~(\ref{WeylOrderedCombinationLaw0}),
\begin{equation}
e^{i k_1 X } e^{ i k_0 T } = e^{ i k_0  T + \left(\frac{k_0/\kappa}{1 - e^{- k_0/\kappa}}\right) k_1 X } \,.
\end{equation}
we get:
\begin{equation}\label{WeylOrderedCombinationLaw}
e^{ i \left( p_0  T + p_1 X \right)} e^{ i \left( q_0  T + q_1 X \right)} 
=
 e^{ i (p_0+q_0)  T + i \left(\frac{(p_0+q_0)/\kappa}{1 - e^{- (p_0+q_0)/\kappa}}\right) \left[ \left(\frac{1 - e^{- \frac{p_0}{\kappa}}}{p_0/\kappa}\right)  p_1 + e^{-\frac{p_0}{\kappa}} \left(\frac{1 - e^{- \frac{q_0}{\kappa}}}{q_0 / \kappa}\right)  q_1 \right] X } 
 \,.
\end{equation}

\subsubsection*{Acknowledgments}
F.L. acknowledges support from the INFN Iniziativa Specifica GeoSymQFT, the Spanish MINECO underProject No. MDM-2014-0369 of ICCUB (Unidad de Excelencia `Maria de Maeztu'), Grant No. FPA2016-76005-C2-1-P.  67985840. F.M. thanks the Action CA18108 QG-MM from the European Cooperation in Science and Technology (COST) and the Foundational Questions Institute (FQXi).

%

\providecommand{\href}[2]{#2}\begingroup\raggedright\endgroup

\end{document}